\documentclass[prd,preprint,amsmath,amssymb,longbibliography]{revtex4-1}
\usepackage{graphicx}
\usepackage{amssymb}
\usepackage{amsmath}
\usepackage[braket, qm]{qcircuit}
\usepackage{bbold}
\usepackage{appendix}

\newcommand{\nmax}{n}
\usepackage{bbm}
\usepackage{lineno}
\begin{document}
\title{Perturbative boundaries of quantum advantage: real-time evolution for digitized $\lambda \phi^4$ lattice models}
\author{Robert Maxton}
\author{Yannick Meurice}
\affiliation{$^1$ Department of Physics and Astronomy, The University of Iowa, Iowa City, IA 52242, USA }

 \date{\today}

\begin{abstract}
The real time evolution of quantum field theory models can be calculated order by order in perturbation theory. For $\lambda \phi^4$ models, the perturbative series have a zero radius of convergence which in part motivated the design of digitized versions suitable for quantum computing. In agreement with general arguments suggesting that a large field cutoff modifies Dyson's reasoning and improves convergence properties, 
we show that the harmonic digitizations of 
$\lambda \phi^4$ lattice field theories lead to weak coupling expansions with a finite radius of convergence. 
Similar convergence properties are found for strong coupling expansions.
We compare the resources needed to calculate the
real-time evolution of the digitized models with perturbative
expansions to those needed to do so with universal quantum computers. Unless new approximate methods can be designed to calculate long perturbative series for large systems efficiently, it appears 
that the use of universal quantum computers with digitizations involving a few qubits per site has the potential for more efficient calculations of the 
real-time evolution for large systems at intermediate  coupling.
\end{abstract}

\maketitle

\def\beq{\begin{equation}}
\def\enq{\end{equation}}
\def\nq{n_q}
\def\nmax{n_{\mathrm{max}}}
\def\phix{\hat{\phi} _{\bf x}}
\def\nq{n_q}
\def\ah{\hat{a}}
\def\ahd{\hat{a}^\dagger}
\def\pn{P_{\nmax-1}}
\def\har{\hat{H}^{\mathrm{har.}}}
\def\hanh{\hat{H}^{\mathrm{anh.}}}
\def\hpf{\hat{\phi}^4}
\def\np{\mathcal{N}_p}

\section{Introduction}
Performing ab-initio real-time calculations for quantum field theory and in particular quantum chromodynamics (QCD) would have a significant impact for theoretical predictions related to  hadron collider experiments \cite{Bauer:2022hpo}. As methods based on probabilistic importance sampling rely on Euclidean time formulations, real-time evolution for lattice QCD cannot be performed with classical computers for 
system sizes comparable to those used for static problems. 
For this reason, the use of universal quantum computers  \cite{lloyd1996universal,Jordan:2011ci,jordan2012quantum,Mezzacapo:2015bra,somma2016quantum,cervera2018exact,Yeter-Aydeniz:2018mix,Raychowdhury:2018osk,Lamm:2018siq,macridin2018digital,klco2018digitization,Bauer:2019qxa,Singh:2019uwd,
Lamm:2019uyc,Lamm:2019bik,Klco:2019evd,Gustafson:2019mpk,Singh:2019uwd,gustafson2021indexed,Gustafson:2021imb,Kharzeev:2020kgc,Honda:2021aum,Barata:2020jtq,Bhattacharya:2020gpm,Ji:2022qvr,Kurkcuoglu:2021dnw,Berenstein:2022wlm,Gustafson:2021jtq,Alam:2021uuq,Caspar:2022llo,Gustafson:2022xdt,Gustafson:2022xlj,Shaw:2020udc,Liu:2021tef,Farrell:2022wyt,Ciavarella:2021nmj,Nguyen:2022aaq,Carena:2022hpz,Carena:2022kpg,Alexandru:2022son,Bauer:2022hpo,Grabowska:2022uos,Murairi:2022zdg}
or analog quantum simulations using cold atoms \cite{bloch2012quantum,lewenstein2007ultracold,cirac2010cold,Kapit:2010qu,Kuno:2016ipi,Martinez:2016yna,Danshita:2016xbo,Zhang:2018ufj,Davoudi:2021ney,Davoudi:2019bhy,Monroe:2019asq,Gonzalez-Cuadra:2017lvz,Nguyen:2021hyk,Aidelsburger:2021mia,Schweizer:2019lwx,meuth}
has become a very active area of research. In this context, 
roadmaps \cite{Dalmonte:2016alw,Banuls:2019bmf,Kasper:2020akk,Bauer:2022hpo,meurice2022tensor} to implement sequences of models of increasing complexity and dimension have been considered, leading to physically relevant calculations on rapidly evolving Noisy Intermediate Scale Quantum (NISQ) platforms \cite{RevModPhys.86.153}.

In one of the first articles discussing real-time evolution and study of scattering processes using universal quantum computers \cite{jordan2012quantum} for $\lambda \phi^4$ theories, it is stated correctly that
traditional calculations of quantum field theory scattering amplitudes rely on perturbation theory and that even at weak coupling, the perturbative series are not convergent. In other words, 
including higher-order contributions beyond a certain
point makes the approximation worse. 

However, in order to set up a quantum computation, finite discretizations of the non-compact scalar field $\phi$ are used 
\cite{jordan2012quantum,somma2016quantum,macridin2018digital,klco2018digitization,Barata:2020jtq,Kurkcuoglu:2021dnw,meurice2021quantum} (we call this process ``digitization"), which implicitly introduces a field cutoff.  It has been argued  \cite{meurice2001simple,kessler} that cutting off the large field contributions in $\lambda \phi^4$ theories affects the instability at negative $\lambda$ invoked by Dyson \cite{dyson} (see Ref. \cite{guillou} for more literature on the subject), and 
results in modified perturbative series that are expected to converge to values exponentially close to exact. Consequently, the question of using perturbative methods in an efficient way needs to be revisited for digitized models. 

Several questions need to be answered for digitized models: 1) do the perturbative series converge? 2) are analytic continuations possible? 3) assuming positive answers for 1) and 2), what are the computational resources needed to perform reasonably accurate calculations? 

In this article we address these questions for lattice versions of $\lambda \phi ^4$ field theory. 
We start with the standard field-continuous formulation of lattice $\lambda\phi^4$ in the local field basis, where $\phix$ is diagonal and the interactions among the fields at different lattice sites $\mathbf{x}$ are limited to 
neighboring sites. For finite local digitizations of these models, general results guarantee \cite{lloyd1996universal}  that quantum computers could deal efficiently with real-time evolution.
Several methods have been used to digitize the local field variables \cite{jordan2012quantum,somma2016quantum,macridin2018digital,klco2018digitization,Barata:2020jtq,Kurkcuoglu:2021dnw,meurice2021quantum}, of which two are discussed here.
In the first, the eigenvalues are equally spaced between $\pm \phi_\mathrm{max}$ for some field cutoff $\phi_\mathrm{max}$ and the conjugate momentum is non-trivial. 
In the other \cite{klco2018digitization,meurice2021quantum}, called the ``harmonic" basis, the field {\it and} its conjugate momentum have eigenvalues which are the 
zeros of the Hermite polynomial $H_{\nmax} (x)$ for some $\nmax$ which denotes the size of the finite local Hilbert space at each lattice site. In the harmonic basis, the standard algebraic manipulations involving creation and annihilation operators hold, with the exception that $a^\dagger \ket{\nmax -1}=0$. 
Because of the simplicity of the conjugate momentum operator and the preservation of most of the algebraic relations involving creation and annihilation operators, the harmonic basis provides easy reformulations of the standard perturbative tools for digitized models and 
will be used almost exclusively hereafter. The construction of the harmonic basis is closely related to the Gaussian quadrature method of integration. With $\nmax$ sampling points, the integration of polynomials of $2\nmax-1$ order remains exact and orthogonality relations are preserved.  It seems clear that by taking $\nmax$ large enough, we recover the integration with continuous and non-compact fields as in 
the standard formulation .

With a universal quantum computer, we use $\nq$ qubits per lattice site and in the following we take $\nmax =2^{\nq}$. 
In view of the limitations of current NISQ devices, we are inclined to consider economical situations with small $\nq$. 
The matching with the target model may not be perfect, however if the symmetries are preserved, we expect that 
in the limit where the lattice is small compared to the correlation length, universal properties of the model will be preserved.
For recent discussions related to this question see e.g., Refs. \cite{Singh:2019uwd,Liu:2021tef,Berenstein:2022wlm,Alam:2021uuq,Caspar:2022llo,Carena:2022hpz,Carena:2022kpg,Alexandru:2022son,Murairi:2022zdg}.

In the following, we discuss the cases $\nq$= 2, 3, 4 and 5 and show that the digitization leads to converging perturbative series, 
unlike the original model with continuous and non-compact fields. For small systems and small couplings, this allows the practical use of perturbation theory to calculate the real-time evolution accurately. The main open question is if this method can be used efficiently when the system size increases and the coupling constant takes arbitrary values.

The article is organized as follows. In Sec. \ref{sec:models}, we present the lattice models considered and their digitization. In Sec. \ref{sec:statement}, we describe the calculation of the matrix elements of the evolution operator in a computational basis using perturbation theory. In Sec. \ref{sec:onesite}, we discuss the one-site problem which is a single digitized quantum anharmonic oscillator. 
In \ref{subsec:methods}, 
we present numerical methods to calculate perturbative series and determine their radius of convergence using the simple case of $\nmax=4$ with one site which can also be solved exactly. 
The complex singularities in the complex $\lambda$ plane for $\nmax$ = 8, 16 and 32 are presented in Sec. \ref{subsec:more}. The complex singularities appear to stay away from the positive real axis. 
We also show that similar methods can be applied in the strong coupling limit. In Sec. \ref{sec:2d}, we consider a 1+1 dimensional model with four sites and $\nmax=4$. 
By increasing the hopping parameter, the complex singularities start pinching the positive real axis in the complex $\lambda$ plane in agreement with the existence of a second-order phase transition. 
Higher dimensional studies are in principle possible but would require optimizations not considered here.
In Sec. \ref{sec:qc}, we discuss the calculation of the same matrix elements with a universal quantum computer. We conclude with a discussion of quantum advantage.
In this context, the recent interest 
\cite{Akiyama:2020ntf,Delcamp:2020hzo,Campos:2019jfa,Vanhecke:2019pez,Vanhecke:2021noi,Campos:2021zce}  in applying tensor network methods to $\lambda \phi^4$ could also provide relevant elements for the discussion. 

\section{Models}
\label{sec:models}
In this section we introduce the lattice models of $\lambda \phi^4$ field theory considered in the article.
These consist of  anharmonic oscillators located at lattice sites with quadratic nearest neighbor coupling.
The target models, before digitization, and the associated terminology are introduced in Sec. \ref{subsec:target}. 
The truncated (digitized) harmonic basis is presented in Sec. \ref{subsec:harmo}. This includes a discussion of the basis where the field operators are diagonal. It should be emphasized that in this ``field basis of the truncated harmonic formulation" the field eigenvalues are not equally spaced. They are the zeros of some Hermite polynomial. This field basis should not be confused with the other field basis mentioned in the introduction, where the field eigenvalues are equally spaced. 

\subsection{Target models: lattice $\phi^4$ Hamiltonians in $D-1$ spatial dimensions}
\label{subsec:target}
In the following we use a spatial lattice. We use the notation 
${\bf x}$ for the lattice sites and ${\bf e}$ for the $D-1$ orthogonal unit vectors in positive directions. 
For instance, for $D=3$ space-time dimensions, this represents a square spatial lattice. The Hamiltonian $\hat{H}$ reads 
\beq
\label{eq:genham}
\hat{H}=\sum_{\bf x}\hat{H}_{\bf x}^{anh.}-2\kappa \sum_{\bf x, e}\hat{\phi}_{\bf x}\hat{\phi}_{\bf x+e},
\enq
with the local anharmonic part 
\beq
\hat{H}_{\bf x}^{anh.}=\omega(\hat{a}_{\bf x}^\dagger\hat{a}_{\bf x} +\mathbf{1}/2)+
\lambda \hat{\phi}_{\bf x}^4.
\enq
Here, $\omega$ represents the energy scale of the unperturbed harmonic oscillators; $\kappa$ represents the hopping energy, the energy scale of the nearest-neighbor coupling; and $\lambda$ determines the strength of the anharmonic component, which in the low-$\lambda$ limit can be interpreted as the strength of the coupling between free particle states. With the 
usual field definition
\beq
\label{eq:phi}
\hat{\phi}_{\bf x}  \equiv \frac{1}{\sqrt{2\omega}}\left(\hat{a}_{\bf x} +\hat{a}_{\bf x}^\dagger\right),
\enq
and the conjugate momentum 
\beq
\hat{\pi}_{\bf x}  \equiv -i \sqrt{\frac{\omega}{2}}\left(\hat{a}_{\bf x} - \hat{a}_{\bf x}^\dagger \right),   
\enq
we have the standard commutation relations
\beq
\left[\hat{a}_{\bf x}, \hat{a}_{\bf y}^\dagger\right] = \delta_{\bf x \bf y}, 
\enq
with the other commutators being zero. 
The above equations provide the standard Hamiltonian formulation of $\lambda \phi^4$.
We will now introduce a truncation of the local Hilbert spaces. 
\subsection{The digitized harmonic basis}
\label{subsec:harmo}

We now consider the harmonic digitization of $\hat{a}_{\bf x}$ and $\hat{a}_{\bf x}^\dagger$ operators at a given site ${\bf x}$. 
As the results are independent of ${\bf x}$ we drop the site index in this subsection.
In order to get a finite-dimensional Hilbert space, we start with the standard 
\beq
\ahd\ket{n}=\sqrt{n+1}\ket{n+1}, {\rm for}\  n=0, \dots, \nmax -2,
\enq
but impose
\beq
\ahd\ket{\nmax -1}=0 .
\enq
In addition, we have the standard relations
\beq
\ah\ket{n}=\sqrt{n}\ket{n-1}, {\rm for}\  n=1, \dots, \nmax -1, 
\enq
and 
\beq
\ah\ket{0}=0.
\enq
This implies that
\beq
\ahd\ah\ket{n}=n\ket{n} , {\rm for}\  n=0, \dots, \nmax -1.
\enq
For this reason we {\it define}   the harmonic Hamiltonian $\har$ as 
\beq 
\har \equiv \omega(\ahd\ah +\frac{\mathbf{1}}{2}),
\enq
which has the same spectrum as the usual harmonic oscillator in the truncated subspace.
Using the notation 
\beq
\pn=
\ket{\nmax -1}\bra{\nmax -1},
\enq
for the projector in the highest energy state,
the modified commutation relations read
\beq
[\ah,\ahd]=\mathbf{1}-\nmax \pn .
\enq
The traces of both sides of the equation are clearly zero. 
Note that 
\beq
\frac{1}{2}({\pi}^2+\omega^2 {\phi }^2)=\har-\frac{\nmax}{2}\pn .
\enq

Since 
\beq\ahd \pn=\pn\ah=0,
\enq
we have the standard relations 
\beq [\ahd \ah,\ah]=-\ah,\ {\rm and}\ [\ahd \ah,\ahd]=\ahd.
\enq
Consequently, we have the standard interaction picture relation
\beq
\label{eq:ai}
e^{i\har t}\  \ah \ e^{-i\har t}={\rm e}^{-i\omega t} \ah\ ,
\enq
and its Hermitian conjugate
\beq
\label{eq:aic}
e^{i\har t}\  \ahd \ e^{-i\har t}={\rm e}^{i\omega t} \ahd .
\enq

\subsection{Local field eigenstates}
\label{subsec:local}

We now discuss the eigenvectors and eigenvalues of the local field $\hat{\phi}$ in the harmonic basis. 
In standard wave mechanics this is called the position basis and the eigenvectors can be expressed in terms of Hermite polynomials. A similar result will be obtained after truncation. The expression of $\hat{\phi}$ in terms of creation and annihilation operators is the standard one given in Eq. (\ref{eq:phi}).

Given a field eigenstate
\beq
\hat{\phi}\ket{\phi}=\phi\ket{\phi},
\enq
and using Eq. (\ref{eq:phi}), we find that, as in the untruncated case, the recursion relation
\beq
\phi\sqrt{2\omega}\langle n |\phi\rangle=\sqrt{n+1}\langle n+1|\phi \rangle+\sqrt{n}\langle n-1|\phi \rangle ,
\enq
can be solved using 
the Hermite recursion relation
\beq 
H_{n+1}(x)-2xH_n(x)+2nH_{n-1}(x)=0\enq
with $x=\sqrt{\omega} \phi$ as in ordinary quantum mechanics \cite{sakurai2011modern}. 
In order to terminate the process at level $n_{max}$ we need to restrict $x$ to values $x_j$ such that
$H_{\nmax}(x_j)=0$. Combining these results with those of Gaussian quadrature integration, we obtain the normalized 
expression
\beq
\langle n |\phi_j\rangle=\sqrt{\frac{2^{\nmax - 1} \nmax ! }{2^n n! \nmax^2 }}\frac{H_n(\sqrt{\omega}\phi_j)}{H_{\nmax -1}(\sqrt{\omega}\phi_j)}.
\enq

\section{Statement of the problem}
\label{sec:statement}

In the following, we consider the real time evolution in a computational basis $\ket{n^{(0)}}$ which corresponds to
the eigenstates of an unperturbed Hamiltonian $\hat{H}_0$:
\beq
\hat{H}_0\ket{n^{(0)}} =E_n^{(0)}\ket{n^{(0)}}. 
\enq
We then consider the real-time evolution operator 
\beq
\hat{U}(t)\equiv e^{-it\hat{H}},
\enq 
for the perturbed Hamiltonian 
\beq
\hat{H}=\hat{H}_0+\lambda \hat{V}. 
\enq
In the case considered here, $\hat{H}_0$ represents the 
quadratic part of the Hamiltonian of Eq. (\ref{eq:genham}), while $\hat{V}$ is the sum of the quartic terms. The identification of the unperturbed basis and the computational basis has also been used in perturbative calculations involving arrays of Rydberg atoms and where the Rabi frequency was treated as a pertubation \cite{meuth}. 

In a typical quantum computation, one prepares the system in an initial state of the computational basis, 
evolve for a time $t$ and then measure the probability to end up in given state of the computational basis. We want to compare the computation of this probability
\beq
Pr\biggl(\ket{n^{(0)}} \rightarrow \ket{m^{(0)}},t\biggr)=\bigg|\bra{m^{(0)}}  \hat{U}(t) \ket{n^{(0)}}\bigg|^2,
\enq
using a quantum computer and perturbation theory. 

The standard methods to calculate the transition probabilities using perturbation theory are reviewed in quantum mechanics textbooks such as Ref. \cite{sakurai2011modern}.
They can be applied to the case of the digitized models. The first one is Dyson's chronological series:
\beq
\bra{m^{(0)}}  \hat{U}(t) \ket{n^{(0)}}=e^{-i E_m^{(0)}t}\bra{m^{(0)}}( \mathbf{1}-i\lambda \int_0^t dt' \hat{V}_I(t')-\lambda^2\int_0^t dt'\int_0^{t'} dt'' \hat{V}_I(t')\hat{V}_I(t'')+ \dots)\ket{n^{(0)}},
\enq
where 
\beq
\hat{V}_I(t)=e^{i \hat{H}_0 t}\hat{V}e^{-i \hat{H}_0 t}.
\enq
As shown in Eqs. (\ref{eq:ai}) and (\ref{eq:aic}), this can be done in the standard way for the digitized oscillator.

The second method consists of inserting the identity formally expressed as the sum of projectors in the perturbed basis
\beq
\label{eq:regpert}
\bra{m^{(0)}}  \hat{U}(t) \ket{m'^{(0)}}=\sum_n\bra{m^{(0)}} n \rangle \langle n \ket{m'^{(0)}}e^{-iE_n t}.
\enq
The perturbed projectors can be calculated using the resolvent \cite{kato2013perturbation}.
\beq
\label{eq:proj}
\ket{n}\bra{n}=\frac{1}{2\pi i}\oint_{\Gamma _n}dz\frac{1}{z-H},
\enq
where $\Gamma_n$ is a complex contour encircling $E_n$ and $E_n^{(0)}$ and no other energy 
level of $\hat{H}$ and $\hat{H}_0$. This can be accomplished if the spectra are nondegenerate and the 
coupling small enough. The perturbative series for the resolvent has the Lippmann-Schwinger 
form
\beq
\label{eq:ls}
\frac{1}{z-\hat{H}}=\frac{1}{z-\hat{H}_0}+\lambda \frac{1}{z-\hat{H}_0}\hat{V}\frac{1}{z-\hat{H}_0}
+\cdots.
\enq
The series expansion for the perturbed energies can be obtained from 
\beq
E_n={\rm Tr}(\hat{H}\ket{n}\bra{n}),
\enq
or just by solving the characteristic equation order by order in $\lambda$.

One should notice that the computational cost of these methods appears to grow exponentially with the perturbative order $\np$. For the calculation of the projector $\ket{n}\bra{n}$ from Eq. (\ref{eq:proj}) and the 
perturbative expansion (\ref{eq:ls}), each factor $1/(z-\hat{H}_0)$ generates a pole in the complex contour when 
acting on $\ket{n}$ and outside the contour otherwise. Consequently, we need to consider $2^{\np}$ cases. For Dyson's series, each $V_I$ factor contains 16 terms (product of creation and annihilation operators) and the apparent cost is $16^{\np}$. 

\section{Numerical methods for one site}
\label{sec:onesite}
In this section, we focus on the one-site problem. 
In Sec. \ref{subsec:methods}, we consider the exactly 
solvable case of $\nmax$ = 4. We present numerical methods and validate them with the exact solution. 
In Sec. \ref{subsec:more}, we present results for 
$\nmax$ = 8, 16 and 32. We also comment about the relation with the strong coupling expansion. 
In this section and the rest of the article, we set $\omega$ = 1. 
\subsection{Exact solutions at $\nmax$ = 4}
\label{subsec:methods}

We first note that $\nmax=2$ is a trivial theory: since the Hamiltonian is invariant under $\phi\to-\phi$, it commutes with the parity operator and thus decouples states of different parities. At $\nmax=2$ there is only one state of each parity and so the Hamiltonian is diagonal and there is nothing to investigate.

For nontrivial cases, the convergence of perturbative expansions in $\lambda$ is limited by exceptional points -- points in the complex $\lambda$ plane where two distinct, coupled energy levels become degenerate \cite{kato2013perturbation}. The nearest such point to the origin then defines the radius of convergence. In the case of $\nmax=4$, the structure of the discretized theory is particularly simple, and a number of relations between the location of the limiting exceptional point and features of the system more accessible to measurement can take a correspondingly simple form.  When we then move to larger operators, the relations so derived no longer give the exact location of the critical lambda, but continue to provide a useful approximation.

In $\nmax=4$, we have
\beq
\ah=\left(
\begin{array}{cccc}
 0 & 1 & 0 & 0 \\
 0 & 0 & \sqrt{2} & 0 \\
 0 & 0 & 0 & \sqrt{3} \\
 0 & 0 & 0 & 0 \\
\end{array}
\right),
\enq
and 
\beq
\label{eq:hanh}
\hanh=\left(
\begin{array}{cccc}
 \frac{3 \lambda }{4}+\frac{1}{2} & 0 & \frac{3 \lambda }{\sqrt{2}} & 0 \\
 0 & \frac{15 \lambda }{4}+\frac{3}{2} & 0 & 3 \sqrt{\frac{3}{2}} \lambda  \\
 \frac{3 \lambda }{\sqrt{2}} & 0 & \frac{27 \lambda }{4}+\frac{5}{2} & 0 \\
 0 & 3 \sqrt{\frac{3}{2}} \lambda  & 0 & \frac{15 \lambda }{4}+\frac{7}{2} \\
\end{array}
\right).
\enq
The characteristic equation reads: 
\begin{eqnarray}
\label{eq:sec} \nonumber
0&= & {\rm det}(\hanh-z\mathbf{1})\\ 
& =&\frac{1}{256}\left(9 \lambda ^2+84 \lambda +16 z^2-120 \lambda  z-48 z+20\right)\nonumber \\
&  &\times \left(9 \lambda
   ^2+300 \lambda +16 z^2-120 \lambda  z-80 z+84\right).
  \end{eqnarray}
 The even levels correspond to the solutions for $z$ for the  first factor and can be expressed as 
 \beq
E_{0,2}= \label{eq:discr}\frac{1}{4} \left(+15 \lambda +6\mp 2 \sqrt{2} \sqrt{27 \lambda ^2+12 \lambda +2}\right).
 \enq
 The odd levels correspond to the solutions for $z$ for the second factor and can be expressed as 
 \beq
E_{1,3}=\frac{1}{4} \left(+15 \lambda +10 \mp2 \sqrt{2} \sqrt{27 \lambda ^2+2}\right).
\enq
These exact solutions can be expanded order by order in $\lambda$ and the radii of convergence are  
the norms of the complex values of $\lambda$ for which the arguments of the square root vanish and consequently, the two eigenvalues become degenerate. 
For the even levels these singularities appear at 
\beq
\lambda=-\frac{1}{9}(2\pm i\sqrt{2}) \simeq -0.222222 - i 0.157135 , 
\label{eq:exactn4}
\enq
and the radius of convergence is $\sqrt{2/27}\simeq 0.272166$. 
For the odd levels the singularities appear at 
\beq
\lambda=\pm i\sqrt{2/27}, 
\enq
and the radius of convergence is also $\sqrt{2/27}$. For larger $\nmax$, odd and even sectors appear to have different radii of convergence. 

Long series for the energy levels can be calculated by constructing a solution of the characteristic equation (\ref{eq:sec}) 
for $z$ 
order by order in $\lambda$. The unperturbed solutions are half-integers and exact arithmetic leads to rational coefficients. For the ground state we have 
\beq
E_0(\nmax=4)=\frac{1}{2}+\frac{3}{4}\lambda -\frac{9}{4}\lambda^2+\frac{27}{4}\lambda^3 +\dots,
\enq
which differs from the usual result \cite{benderwu}
\beq
E_0(\nmax=\infty)=\frac{1}{2}+\frac{3}{4}\lambda -\frac{21}{8}\lambda^2+\frac{333}{16}\lambda^3 +\dots,
\enq
at order $\lambda^2$. This can be explained from the fact that for $\nmax=4$, $\bra{0^{(0)}} \hat{V} \ket{4^{(0)}} =0$ instead of $\sqrt{6}/2$ in the untruncated case, which modifies the contributions from 
the standard perturbative formula at second order. We will now see that the truncation modifies the asymptotic behavior of the series.  

\begin{figure}[h]
\includegraphics[width=8.cm]{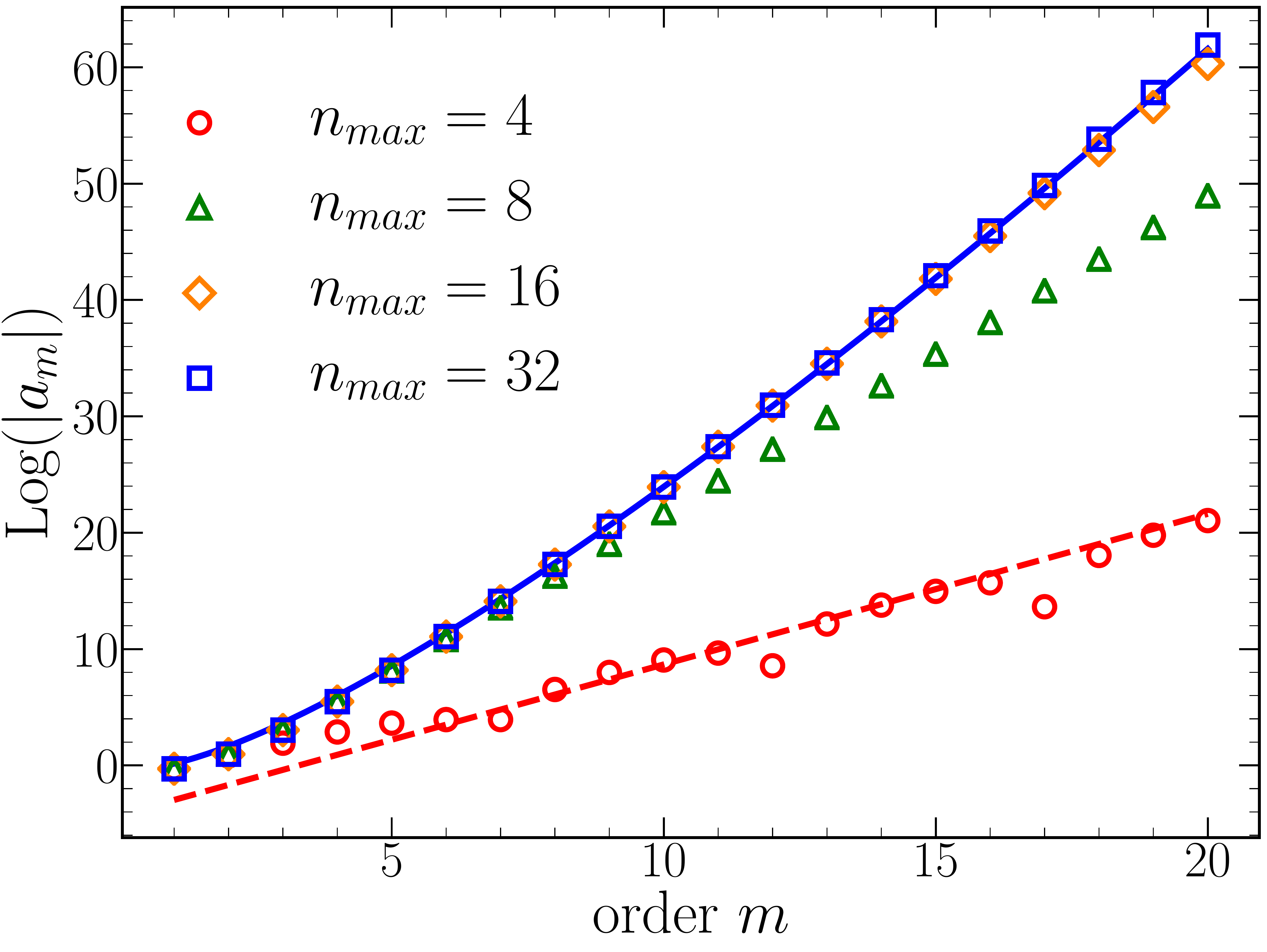} 
\caption{\label{fig:bw} Logarithms of the absolute value of the perturbative coefficients for the ground state of the anharmonic oscillator for $\nmax$= 4 (circles), 8 (triangles), 16 (diamonds)  and 32 (squares). The solid line is the Bender-Wu formula Eq. (\ref{eq:bw}).
The dashed line has a slope $\log(\sqrt{27/2})$.}
\end{figure}

Our calculation of the expansion coefficients can be verified by comparing the our results at larger $\nmax$ (discussed in Sec. \ref{subsec:more}) with the well-established results in the continuous field limit \cite{benderwu}. The comparison between the coefficients for $\nmax$= 4, 8, 16  and 32 is given in Fig. \ref{fig:bw}.  The coefficients for $\nmax=32$ agree with the tabulated results of Ref. \cite{benderwu} to at least 5 significant digits and can be compared with the asymptotic formula, found by the same authors, for the coefficients $a_m$ at order $m$ \cite{benderwu}:
\beq
\label{eq:bw}
|a_m (\nmax=\infty)
| \sim\sqrt{\frac{6}{\pi^3}}3^m\Gamma(m + 1/2).
\enq
On the other hand, from the location of the singularities for $\nmax = 4$ where the argument of the square root vanishes, we expect
\beq
|a_m \big(\nmax=4)|\propto( \sqrt{27/2}\big)^m,
\enq 
which implies an approximate slope of $\log(\sqrt{27/2})\simeq$ 1.30134 for the logarithm of the coefficients observed in Fig. \ref{fig:bw}. The departure from the linear behavior comes from the imaginary part of the singularities and $1/m$ corrections. As the series is calculated up to large order in $\lambda$, these corrections average out or disappear. For instance, by fitting the coefficients between order 50 and 100, we obtain a slope 1.28493, while using orders between 100 and 200, we obtain a slope 1.29227.
The coefficients for larger $\nmax$ have larger slopes which will be discussed in Sec. \ref{subsec:more}.

\begin{figure}[h]
\includegraphics[width=8.6cm]{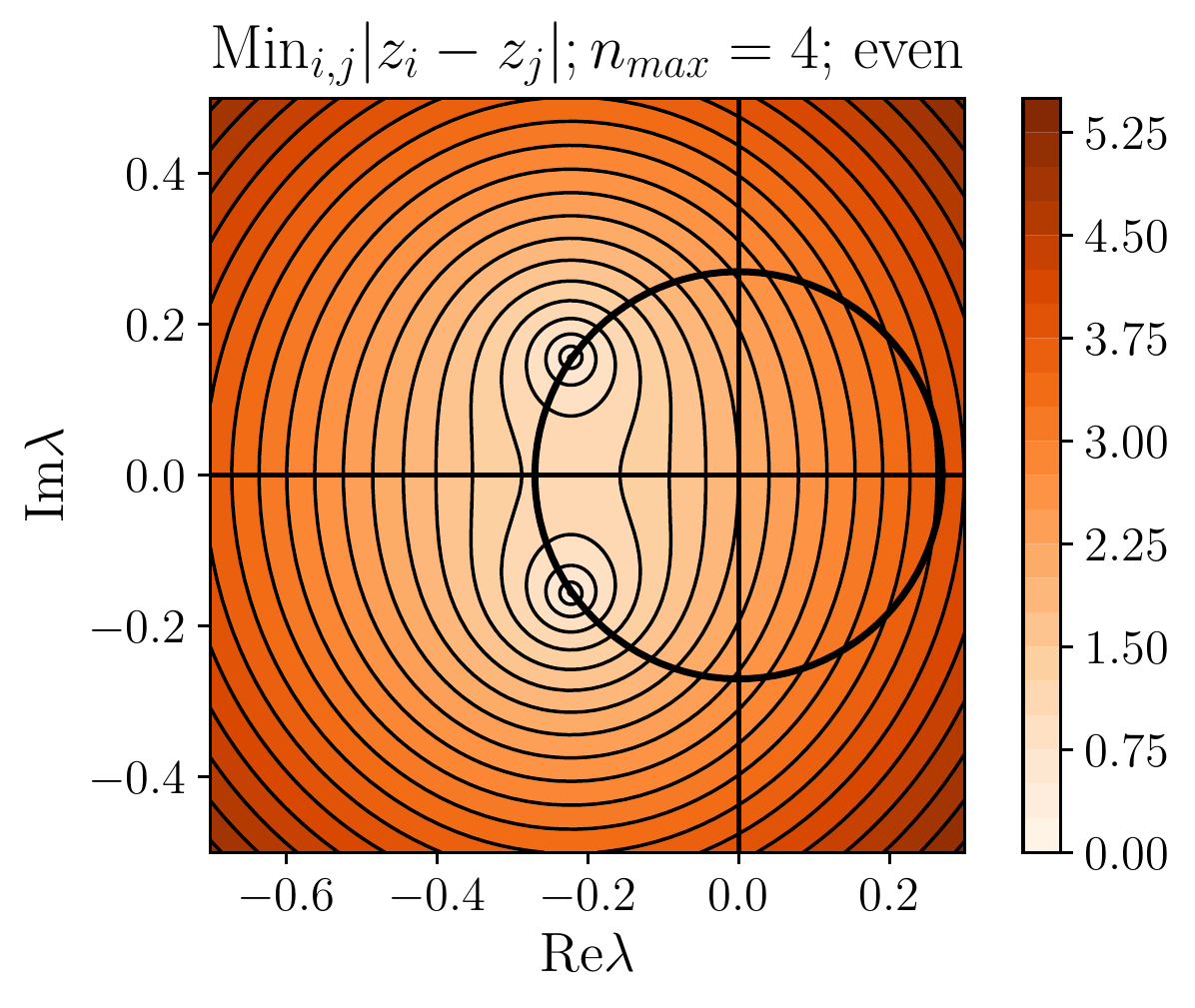} 
\caption{\label{fig:deg4} Minimum of $|z_i-z_j|$ for every possible pairs $i,j$ of even $H$ eigenvalues in the 
complex $\lambda$ plane for $\nmax$ = 4. The zeros appear near the values $-0.222 \pm i 0.157$ expected from the exact answer from Eq. (\ref{eq:exactn4}). The dark circle around the origin represents the boundary of the disk of convergence of the ground state. }
\end{figure}

There are simpler ways to estimate the radius of convergence and the corresponding complex singularities. The first one is to study the second derivative of the energy with respect to $\lambda$ 
on the real axis which has a peak at the real part of the complex singularity and a width at half maximum $W_{1/2}$ which is related to imaginary part of the singular value $\lambda_s$ by the simple relation
\beq
\label{eq:hw}
W_{1/2}=(2\sqrt{2^{2/3}-1}) {\rm Im}\lambda_s,
\enq
as can be shown from the second derivative of the exact expression in Eq. (\ref{eq:discr}). Alternatively, the width is encoded into the magnitude of the fourth derivative, so that we can instead use
\beq
\label{eq:imag}
{\rm Im}\lambda_s = \sqrt{-3\frac{E^{\prime\prime}({\rm Re}\lambda_s)}{E^{(4)}({\rm Re}\lambda_s)}}.
\enq 
Another method is to directly search for degenerate eigenvalues of $H$ in the complex $\lambda$ plane. This is accomplished by calculating the eigenvalues $z_i$ of $H$ for a fine grid in the complex $\lambda$ plane, and displaying the minimum of $|z_i-z_j|$ for every possible pairs $i,j$ of roots in a relevant region. 
The results are illustrated in Fig. \ref{fig:deg4}.

\begin{figure}[h]
\includegraphics[width=8.cm]{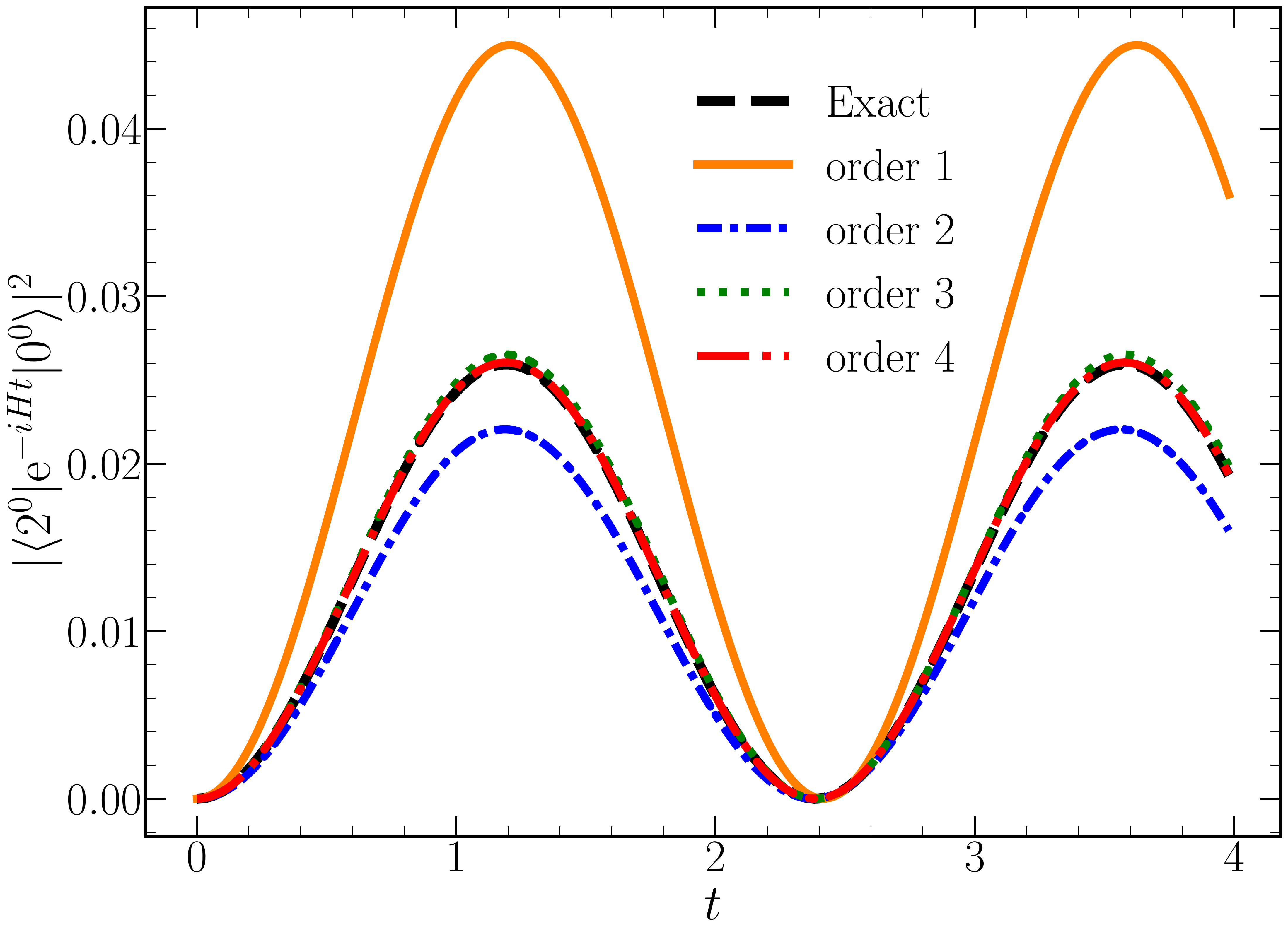} 
\caption{\label{fig:succ} Four successive approximations of $P(0\rightarrow 2,t)$ for $\lambda=0.1$.  }
\end{figure}

From the above discussion, we expect that it is possible to construct converging weak series for the evolution operator when 
$|\lambda|<\sqrt{2/27}\simeq 0.272166$.  As an illustration of convergence, we have calculated 
the probability for an initial state $\ket{0^0}$ to transition to $\ket{2^0}$ after a time $t$:
\beq
P(0\rightarrow 2,t)\equiv |\bra{2^0}\hat{U}(t)\ket{0^0}|^2,
\enq
at four successive orders of $\lambda=0.1$ using the projectors described in Eq. (\ref{eq:proj}). 
The details of the calculation are provided in Appendix \ref{app:projectors}. 
The numerical results are shown in Fig. \ref{fig:succ}. The results at fourth order are difficult to distinguish from the accurate numerical values obtained from the exact result.

\begin{figure}[h]
\includegraphics[width=8.cm]{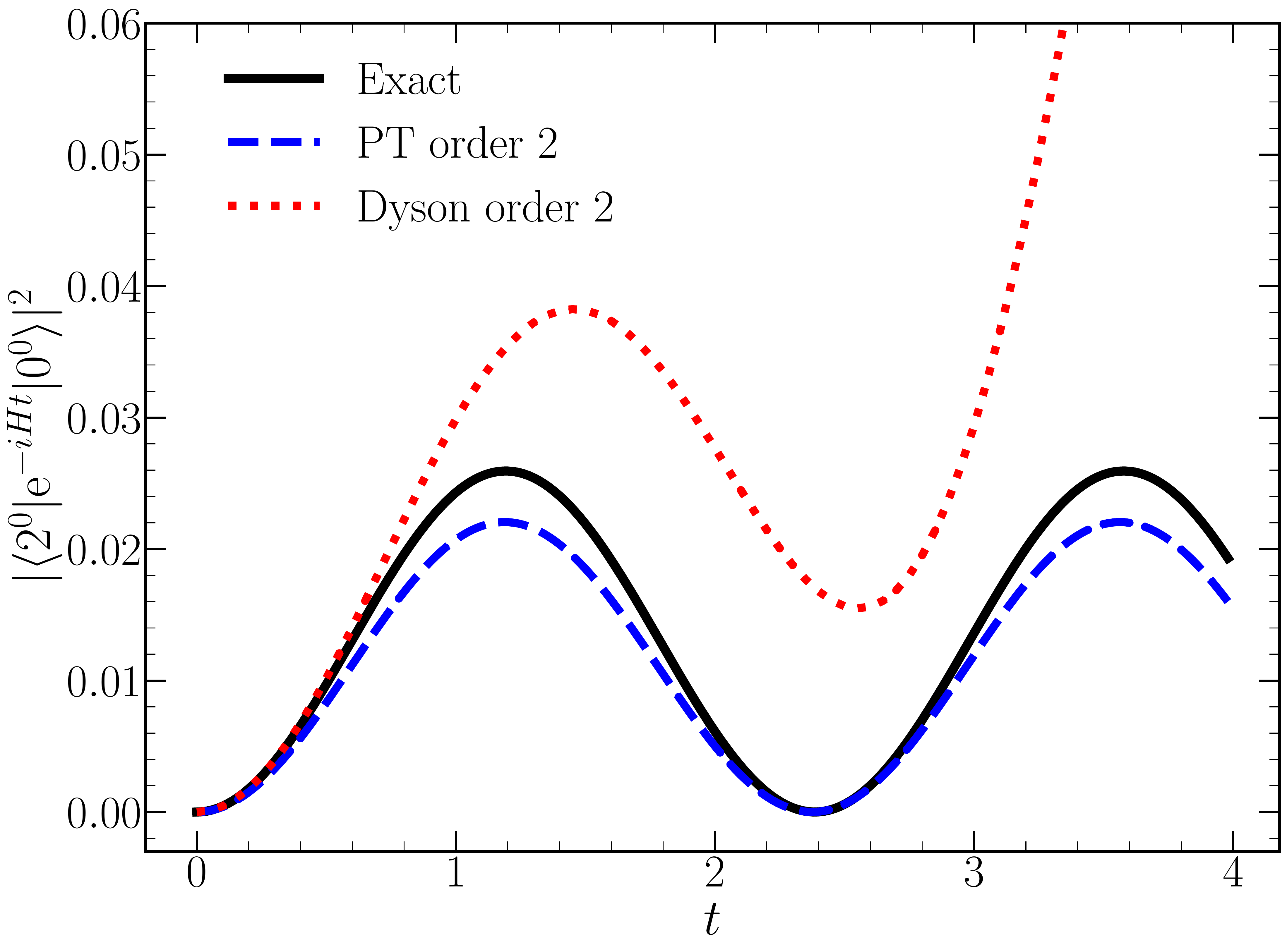} 
\caption{\label{fig:dyson} Comparison of $P(0\rightarrow 2,t)$ for $\lambda=0.1$ calculated at order 2 using Eq. (\ref{eq:regpert}) and  Dyson series.}
\end{figure}

The Dyson series can be calculated independently and checked by expanding the evolution operator calculated above using Eq. (\ref{eq:regpert}). For the 0 to 2 transition, the result up to order 2 is
\beq
\bra{2^0}\hat{U}(t)\ket{0^0}=-\frac{3 i \lambda  e^{i t} \sin (t)}{\sqrt{2}} +\frac{9 \lambda ^2 \left(i e^{2 i t} t-9 i t+4 e^{2 i t}-4\right)}{8 \sqrt{2}}+\mathcal{O}(\lambda^3).
   \enq
The expression contains polynomials in $t$ that for large enough $t$ become poor approximations of the exponentials. 
This is illustrated in Fig. \ref{fig:dyson} at order two. 
In contrast with the approximate solutions based on Eq. (\ref{eq:regpert}) in Fig. \ref{fig:succ}, the approximate energies are inserted in $e^{-iE_n t}$ and the phase factors remain periodic and bounded for any $t$.

Assuming that for $\lambda$ small enough, all the (time-independent) perturbed energies and matrix elements of the perturbed energy projector converge, does Dyson's series converge for arbitrary $t$? Given that we only have a finite number of states, this amounts to ask: if $\lambda$ is small enough and $E_n=\sum_{m=0}^{\infty}\lambda ^mE_n^{(m)}$ converges to the correct energy, 
does $e^{-itE_n}\simeq\sum_{m=0}^{m_{max}}\lambda ^m b_m(t)$ converge to the correct phase for any $t$?
The analyticity of an energy eigenvalue implies the analyticity of its exponential, so the operator series will converge for the smallest radius of convergence of its eigenvalues. 
 Empirically, for $\nmax=4$, $\lambda=0.1$, there is an apparent convergence for $t$ at least up to 100.
For $\lambda=0.3>0.272 ..$, the series clearly diverges for any t. Finding rigorous bounds on Dyson series directly from its series expansion seems to be non-trivial \cite{moan}.

It is also possible to set up a strong coupling expansion for the problems discussed above. 
Using $\hpf$ as the unperturbed Hamiltonian we can define 
\beq
\hat{H}^{\textrm{str.}}=\hpf+\tilde{\lambda}\har=\hanh / \lambda ,
\enq
with $\tilde{\lambda}=1/\lambda$. 
It is clear that for a finite and nonzero $\lambda$, a solution $z(\lambda)$ of the characteristic equation for 
$\hanh$  gives a solution $\tilde{z}(\tilde{\lambda})=z(\lambda)/\lambda$ of the characteristic equation for $\hat{H}^{\textrm{str.}}$.

Again, $\nmax=4$ makes everything simple. Looking at Eq. \ref{eq:discr}, for example, we see that by simply pulling a factor of $\lambda$ out in front of the parentheses, we get immediately

\beq
\lambda E_{0,2}^{\textrm{str.}} = \frac{\lambda}{4} \left(15 + 6\lambda^{-1} \mp 2\sqrt{2} \sqrt{27 + 12\lambda^{-1} + 2\lambda^{-2}}\right).
\enq
Setting the polynomial under the root to zero and solving for $\lambda^{-1}$ immediately gives precisely the inverse of our previous result for $\lambda$. 

This result can also be understood geometrically. The radius of convergence is just the distance between the origin point of the expansion and the exceptional point nearest to it. Weak coupling picks $0$ as its origin; strong coupling picks $\infty$. But at $\nmax=4$, there are only two exceptional points in the first place, which are equidistant from the origin; so starting at zero or complex infinity, the boundary line ends up in the same place.

\subsection{\label{subsec:more}$\nmax$= 8, 16 and 32}

\begin{figure}[h]
\includegraphics[width=8.cm]{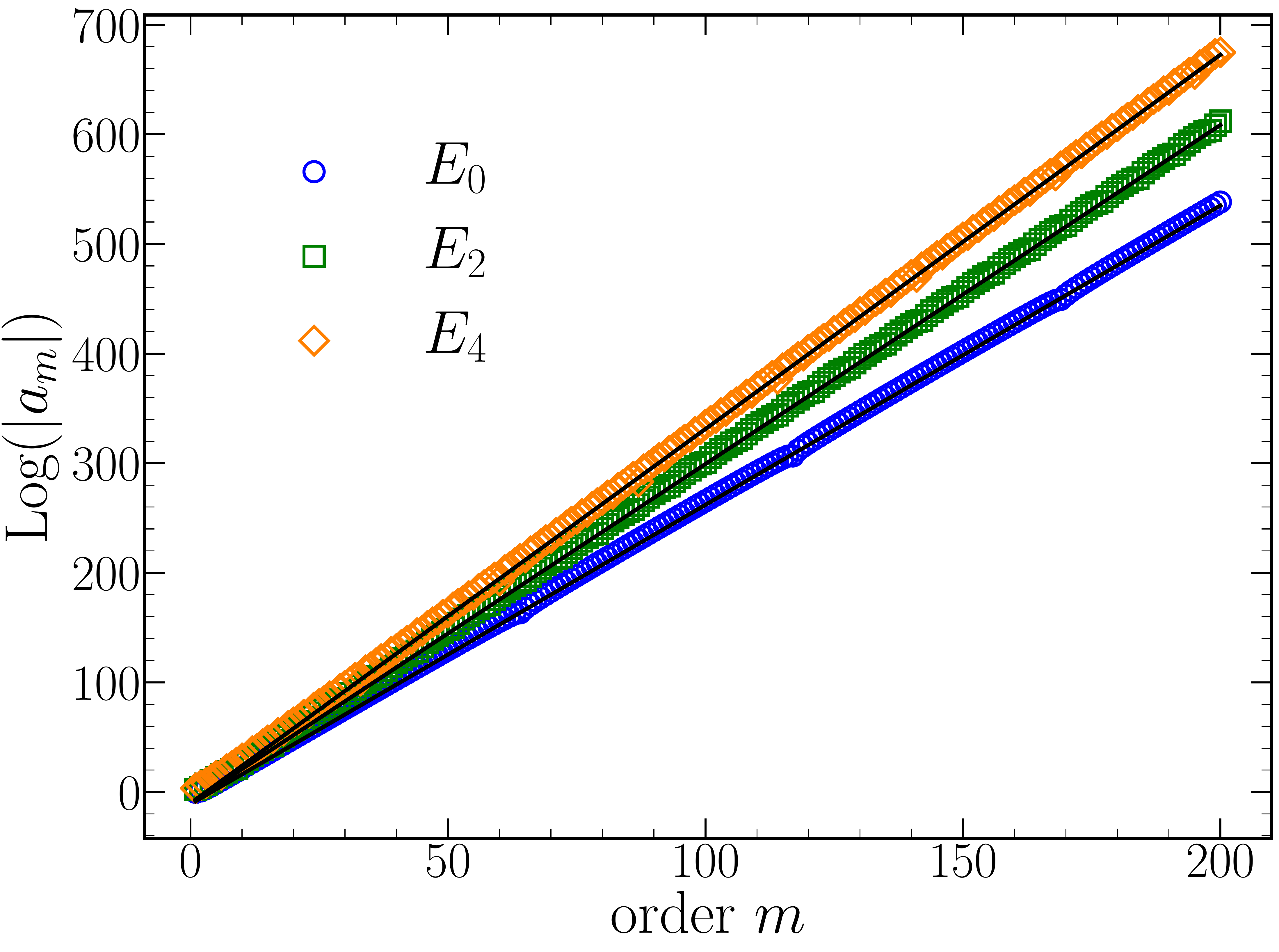} 
\caption{\label{fig:pslopes} Logarithms of the absolute value of the perturbative coefficients for the ground state (circles), second excited state (squares) and fourth exited state (diamonds) of the anharmonic oscillator for $\nmax$= 8. The straight lines represent linear fits of the last hundred coefficients. The estimates of the radius of convergence given in the middle column of Table \ref{tab:rad} is the exponential of minus these slopes.}
\end{figure}

All the numerical methods developed for $\nmax$ = 4 can be used for larger values of $\nmax$. 
 For instance, the logarithms of the absolute values of the coefficients also show a linear behavior. 
The finite radii of convergence can be estimated using a linear fit of the asymptotic behavior. 
This is illustrated for the even sector for $\nmax$ = 8 in Fig. \ref{fig:pslopes}. For a linear fit $a+bm$ with $m$ the order, the estimate for the radius of convergence is $\exp(-b)$.
The estimates based on linear fits between 100 and 200 
are given in Table \ref{tab:rad}. The estimate for the radius for $E_6$ is the same as for $E_4$. 

\begin{figure}[h]
\includegraphics[width=8.cm]{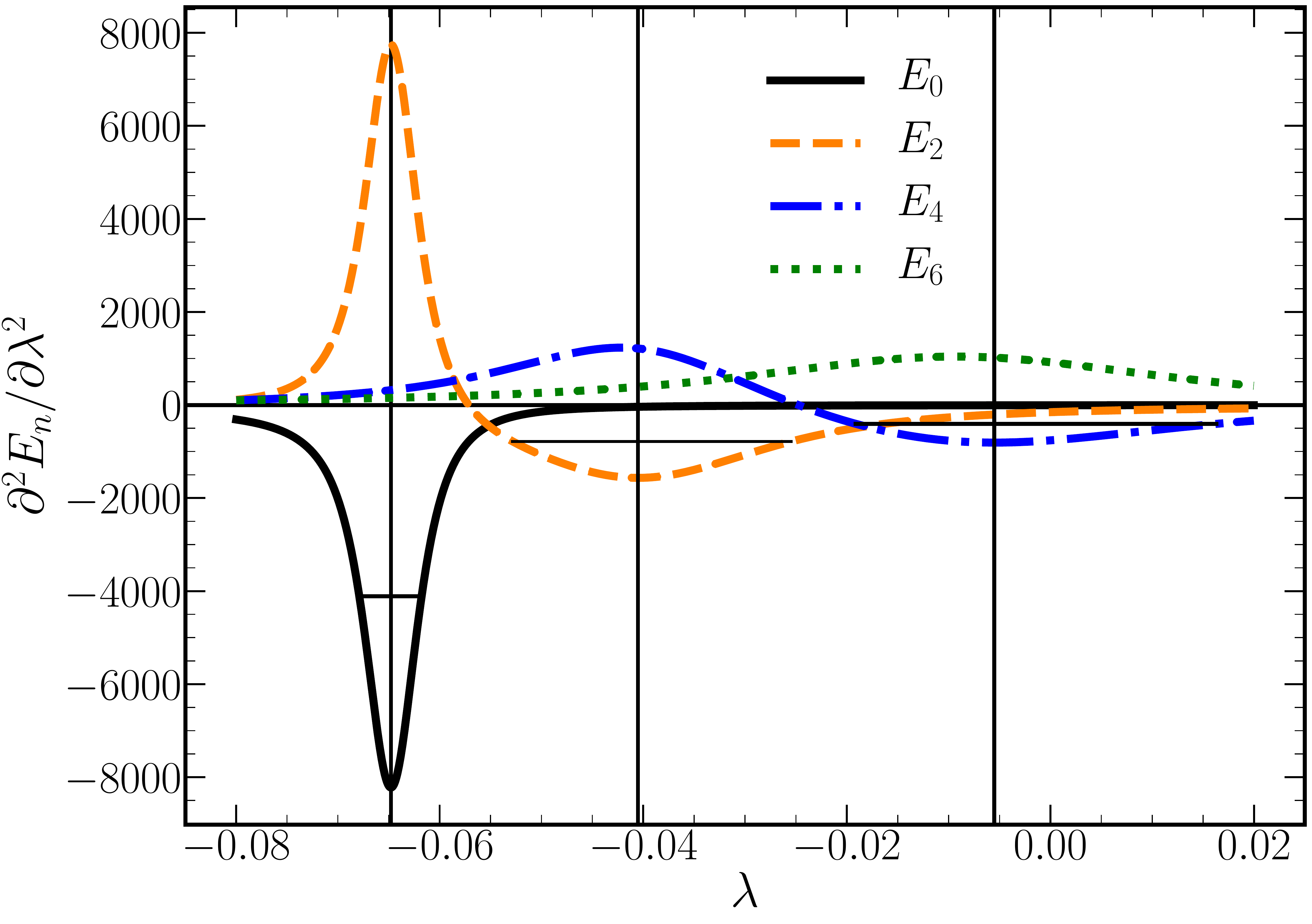} 
\caption{\label{fig:secder}Second derivatives of $E_0$, $E_2$, $E_4$ and $E_6$ with respect to $\lambda$ on the real axis  for $\nmax=$8. The vertical lines are located at the minima and the horizontal lines represent the width at half maximum. }
\end{figure}

\begin{figure}[h]
\includegraphics[width=8.6cm]{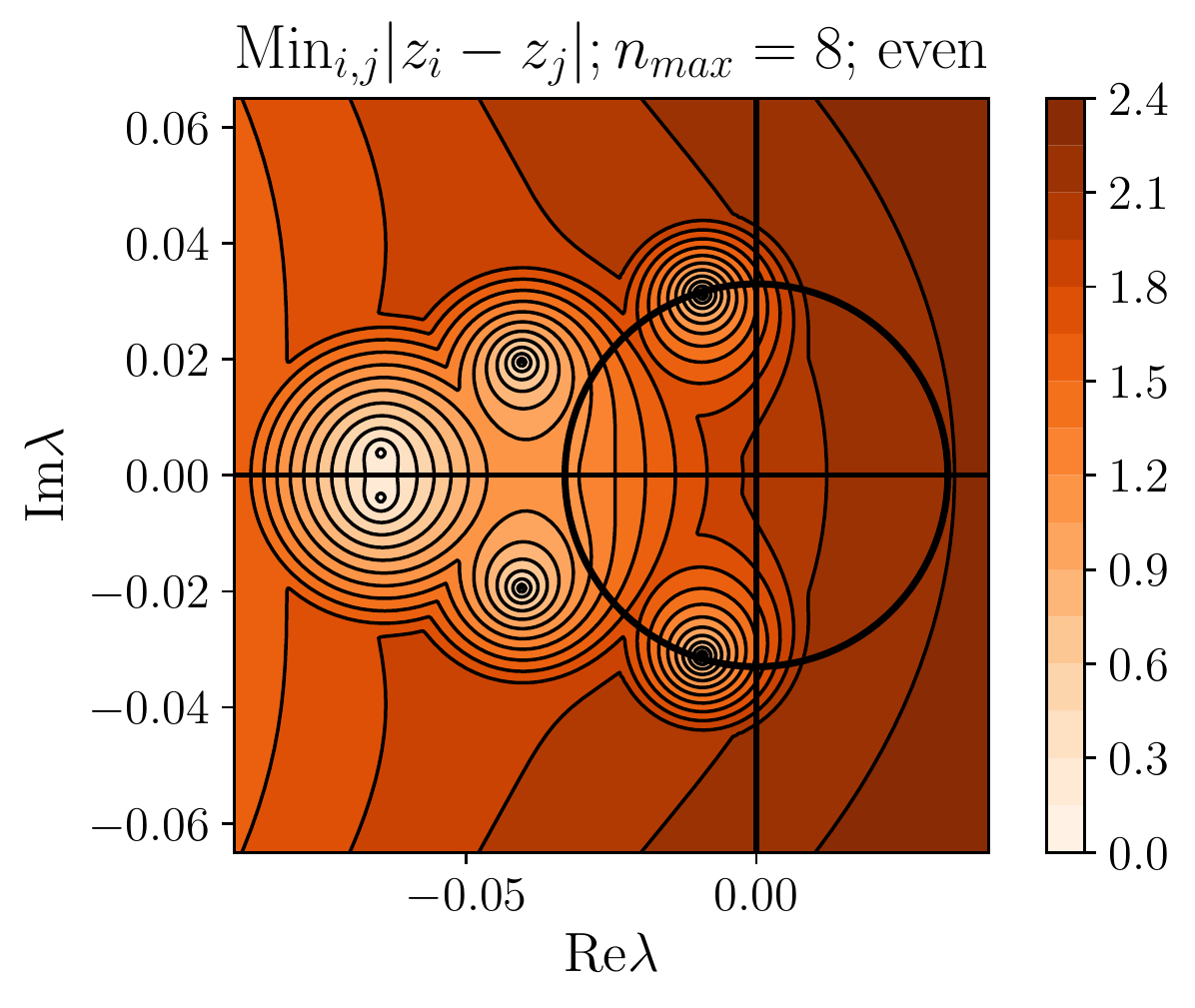} 
\caption{\label{fig:deg8even} Minimum of $|z_i-z_j|$ for every possible pair $i,j$ of even $H$ eigenvalues  for $\nmax$=8 in the 
complex $\lambda$ plane. The dark circle around the origin represents the circle of convergence of the fourth and sixth excited states and has radius 0.033.}
\end{figure}

It is also possible to get reasonable estimates of the radius of convergence by studying the second derivative of the energy.  As there are now several singularities, the relation between the width-at-half-max of the peak of the second derivative and the imaginary component of the radius of convergence no longer exactly obeys Eq. \eqref{eq:hw}, which was derived in the simpler context of $\nmax$ = 4 where the spectrum can be obtained in closed form by solving quadratic equations. However, Eq. \eqref{eq:hw} continues to provide good estimates and a simple picture that the singularities come in pairs: 0-2, 2-4 and 4-6, and the pairs share extrema of $\partial ^2 E/\partial \lambda ^2$. This is illustrated in Fig. \ref{fig:secder}. As we consider higher energy levels, the curve broadens and the imaginary component becomes important. For the degeneracy between $E_4$ and $E_6$, the direct search in Fig. \ref{fig:deg8even} 
shows that the estimate 0.033 of the radius of convergence from the series is accurate. However, when the peak is narrow, more accurate estimates can be obtained. For instance, for the degeneracy between $E_0$ and $E_2$, 
the second derivative provides a value of -0.0648 for the real part and 0.00399 for the imaginary part which are in reasonable agreement with the more accurate values -0.06473 and 0.00391 obtained by direct search.
\begin{table}[h]
 \begin{tabular}{|c|c|c|}
\hline
Level $n$& $|\lambda_s|$ (series)&$|\lambda_s|$ (der.)\\
\hline
0&0.0651&0.0649\\
2&0.0454&0.0456\\
4&0.0329&0.0330\\
\hline
\end{tabular}
\caption{\label{tab:rad} Estimates of the radius of convergence for $E_0$, $E_2$ and $E_4$ using the perturbative series and the second derivative methods for $\nmax$=8.}
\end{table}

\begin{figure}[h]
\includegraphics[width=8.6cm]{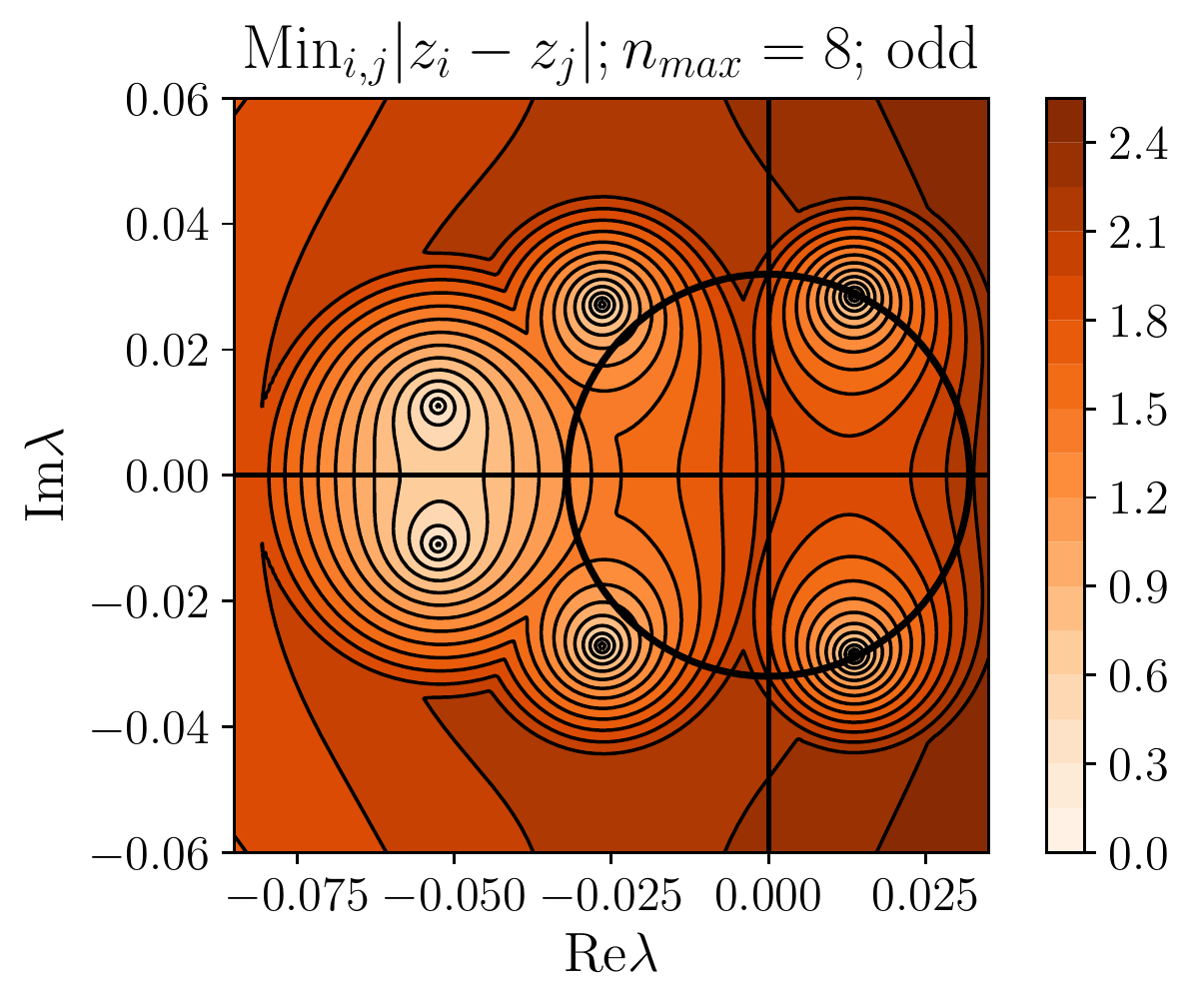} 
\caption{\label{fig:deg8odd} Minimum of $|z_i-z_j|$ for every possible pair $i,j$ of odd $H$ eigenvalues  for $\nmax$=8 in the 
complex $\lambda$ plane. The dark circle around the origin represents the circle of convergence of the fifth and seventh excited states and has an approximate radius of 0.032.}
\end{figure}

The direct search for degeneracies in the complex $\lambda$ plane provides a more holistic view of the position of the singularities \cite{kato2013perturbation}. Fig. \ref{fig:deg8even}  shows that within a circle of radius 0.033, the expansions for all the energy levels should converge. Furthermore, everything we have done in the even sector can be repeated for the odd sector. The results are shown in Fig. \ref{fig:deg8odd} and obey the same qualitative pattern. 

\begin{figure}[h]
\includegraphics[width=8.6cm]{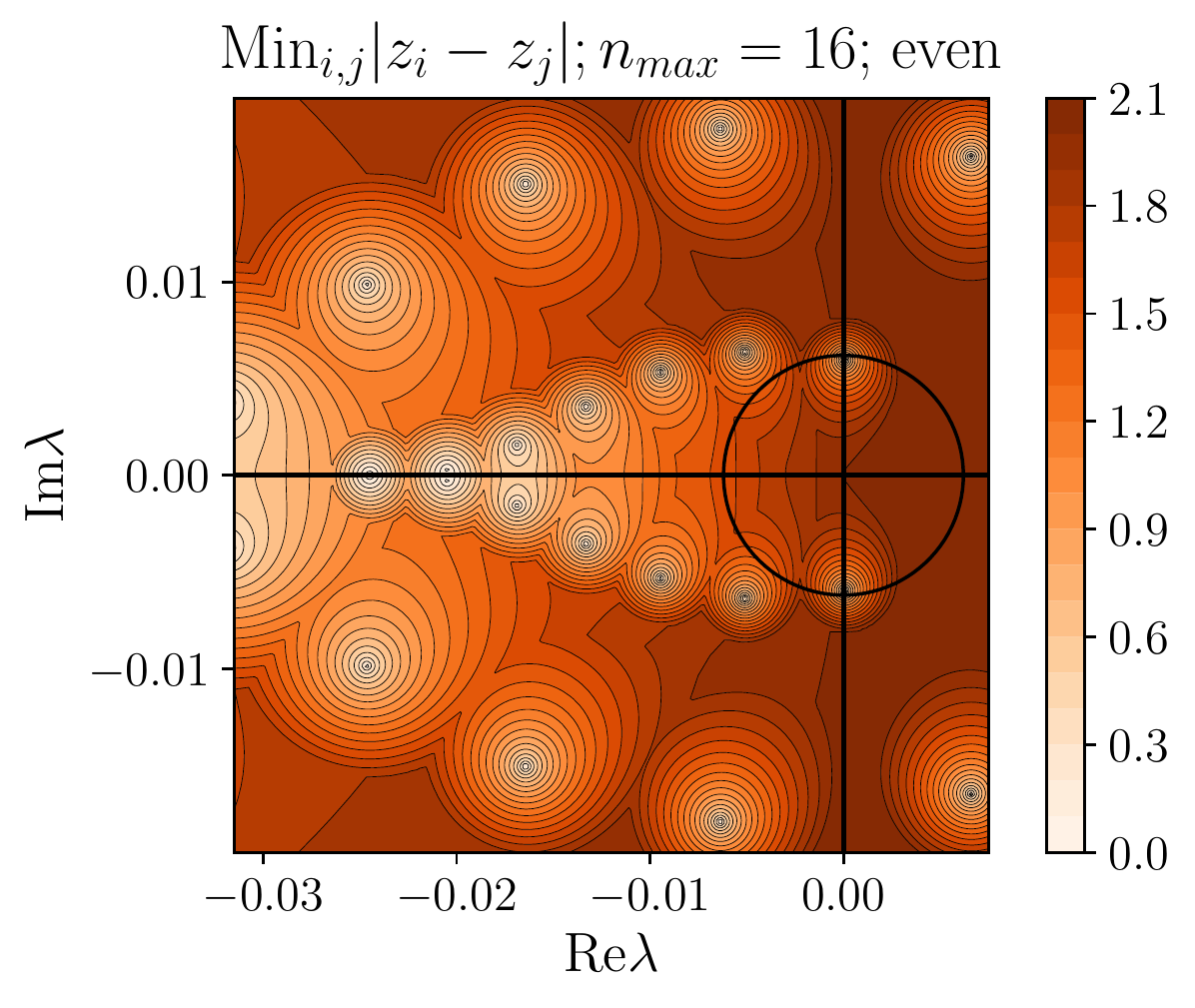} 
\includegraphics[width=8.6cm]{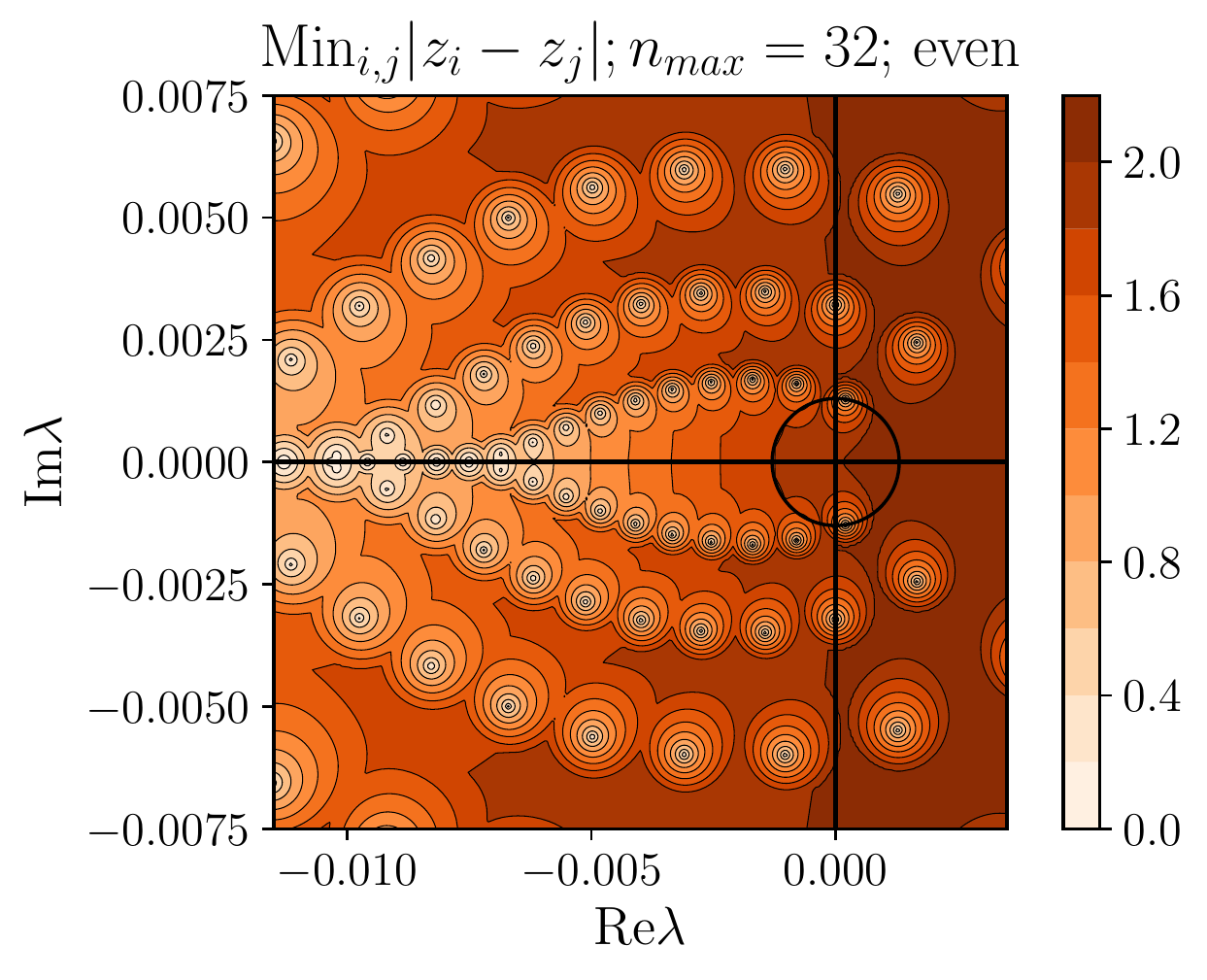} 
\caption{\label{fig:deg16even} Minimum of $|z_i-z_j|$ for every possible pair $i,j$ of even $H$ eigenvalues  for $\nmax$=16 (top) and $\nmax$=32 (bottom) in the 
complex $\lambda$ plane. The dark circle around the origin represents the circle of convergence of the 12th and 14th excited states for $\nmax$=16 and the radius is approximately 0.0060.  For $\nmax$=32, the dark circle represents the circle of convergence of the 28th and 30th excited states and the radius is approximately 0.0013.}
\end{figure}

The direct search is also possible for $\nmax$ = 16 and 32. The results are shown in Fig. \ref{fig:deg16even}. Additional layers appear but the patterns closest to the origin can be recognized. For $\nmax=16$, the fits of perturbative coefficients between 50 and 100 provide estimates of the radii of convergence given in Table \ref{tab:rad16}. 
\begin{table}[h]
 \begin{tabular}{|c|c|}
\hline
Level $n$& $|\lambda_s|$ (series)\\
\hline
0 & 0.0245 \\
 2 & 0.0205 \\
 4 & 0.0191 \\
 6 & 0.0144 \\
 8 & 0.0113 \\
 10 & 0.00864 \\
 12 & 0.00621 \\
 14 & 0.00621 \\
\hline
\end{tabular}
\caption{\label{tab:rad16} Estimates of the radii of convergence for $E_0$ to $E_{14}$ at $\nmax=16$ using the perturbative series. }
\end{table}

They approximately correspond to the seven pairs of complex conjugate singularities observed in Fig. \ref{fig:deg16even}.
Note that for the ground state, the imaginary parts are very small ($\pm 0.000015$) and cannot be resolved in the figure. For $\nmax=32$, the fits of perturbative coefficients between 50 and 100 provide estimates of the radii of convergence which are between 0.0103 for the ground state and 
0.00132 for the 28th and 30th excited states which also seem consistent with Fig. \ref{fig:deg16even}. Note that as $\nmax$ increases, the radius of convergence of the perturbative series decreases approximately as $C/\nmax$ for some constant $C$. This observation can be justified under the assumption that there is a rather sudden transition at negative $\lambda$ when the harmonic and anharmonic terms have the same magnitude but opposite signs for the largest  eigenvalue of the field $x_{max}$ (which scales like $\nmax^{1/2}$). In other words, the transition occurs when $|\lambda| x_{max}^4\simeq (1/2) x_{max}^2$, and we we expect $|\lambda_c| \simeq 1/(2 x_{max}^2) \propto 1/\nmax$.

Almost everything done for the weak coupling expansion can be done for the strong coupling expansion. The eigenstates of the unperturbed Hamiltonian are the  field eigenstates discussed in Sec. \ref{subsec:local}.
For $\nmax$ even, all the unperturbed eigenstates of $\hpf$ are doubly degenerate, however if we project into the 
independent odd and even sectors of the Hilbert space, the degeneracy disappears in each sector. 
The eigenvalues of the unperturbed Hamiltonian are the fourth power of the zeros of the Hermite polynomial of degree $\nmax$ and they need to be calculated numerically. The recursive calculation of the perturbative coefficients cannot efficiently be done by exact arithmetic (due to the presence of products of long summations) and one needs to control the error propagation by using an appropriate working precision. Linear fits can be performed, and the second derivatives can be calculated as before.

Again the main results can be summarized using the minimum of $|\tilde{z}_i-\tilde{z}_j|$ to identify  the singular points and estimate the radii of convergence. The results near $\tilde{\lambda}=0$ for $\nmax=8$ are shown below in Fig. \ref{fig:deg8evenSC}. 
They are qualitatively similar to the results at weak coupling. If we extend the range, we find the singular points already identified at weak coupling. The quantitative correspondence is discussed in Appendix \ref{app:stw}. Note also that the questions of singularities can also be investigated by purely algebraic methods \cite{hao}. This is discussed in Appendix \ref{app:algebra}.

\begin{figure}[h]
\includegraphics[width=8.cm]{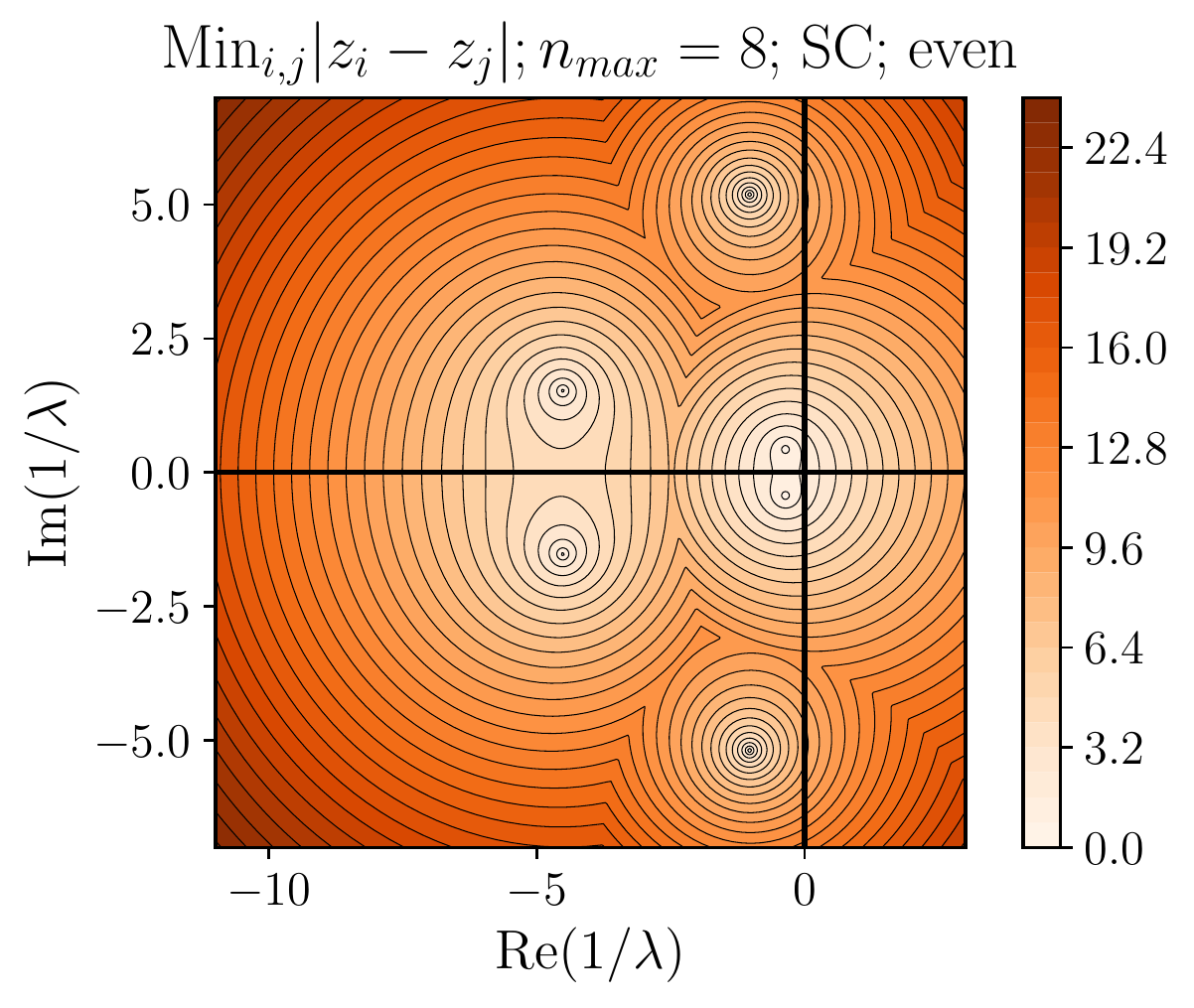} 
\caption{\label{fig:deg8evenSC} Minimum of $|\tilde{z}_i-\tilde{z}_j|$ for every possible pair $i,j$ of even $H^{str.}$ eigenvalues  for $\nmax$=8 in the 
complex $1/\lambda$ plane. }
\end{figure}

\begin{figure}[h]
\includegraphics[width=8.6cm]{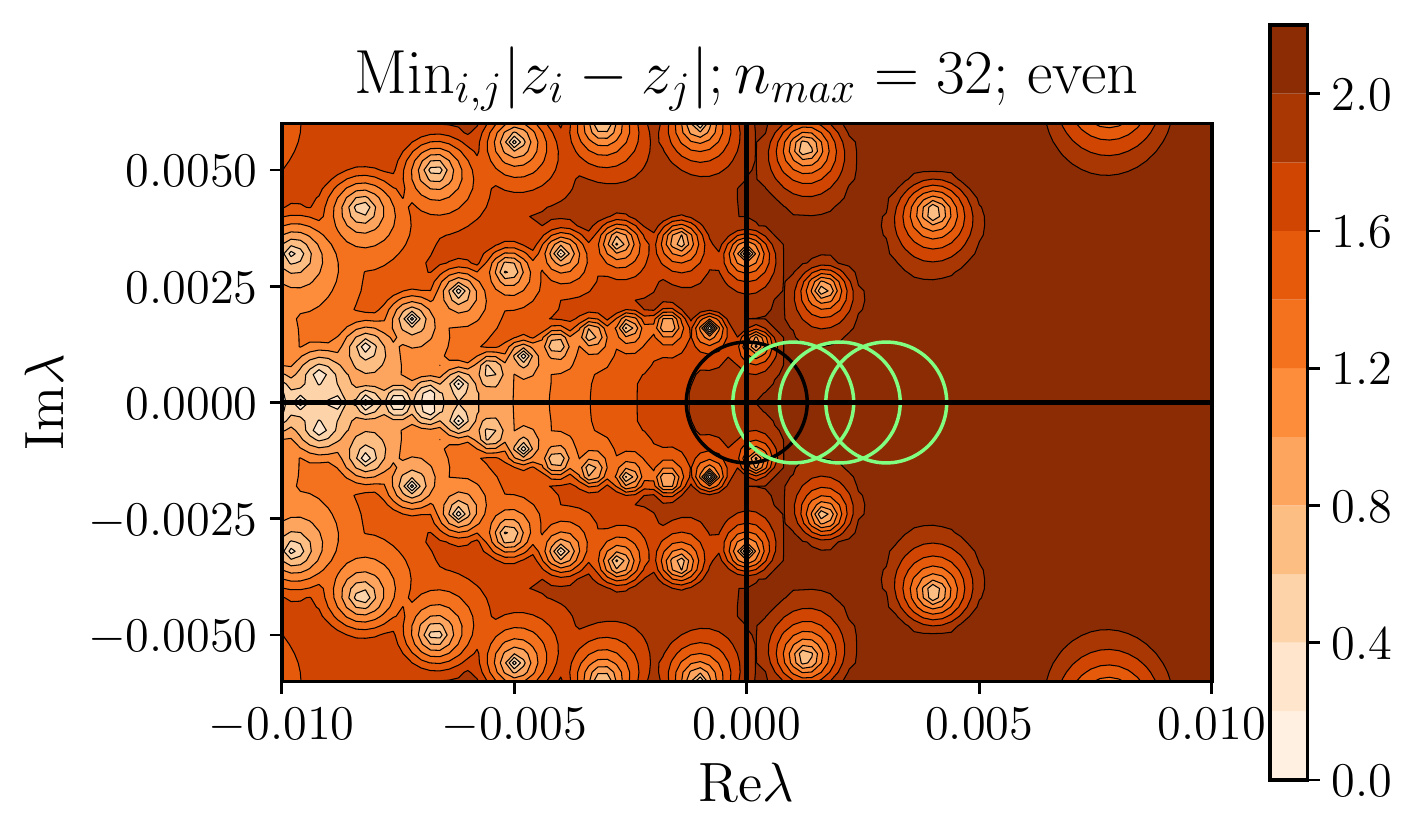} 
\caption{\label{fig:deg32evencont} Minimum of $|z_i-z_j|$ for every possible pair $i,j$ of even $H$ eigenvalues  for $\nmax$=32 in the 
complex $\lambda$ plane. The dark circle around the origin represents the circle of convergence of the 28th and 30th excited states and the radius is approximately 0.0013. The bright circles with the same radius illustrate a possible route for analytic continuation.}
\end{figure}

\begin{figure}[h]
\includegraphics[width=8.6cm]{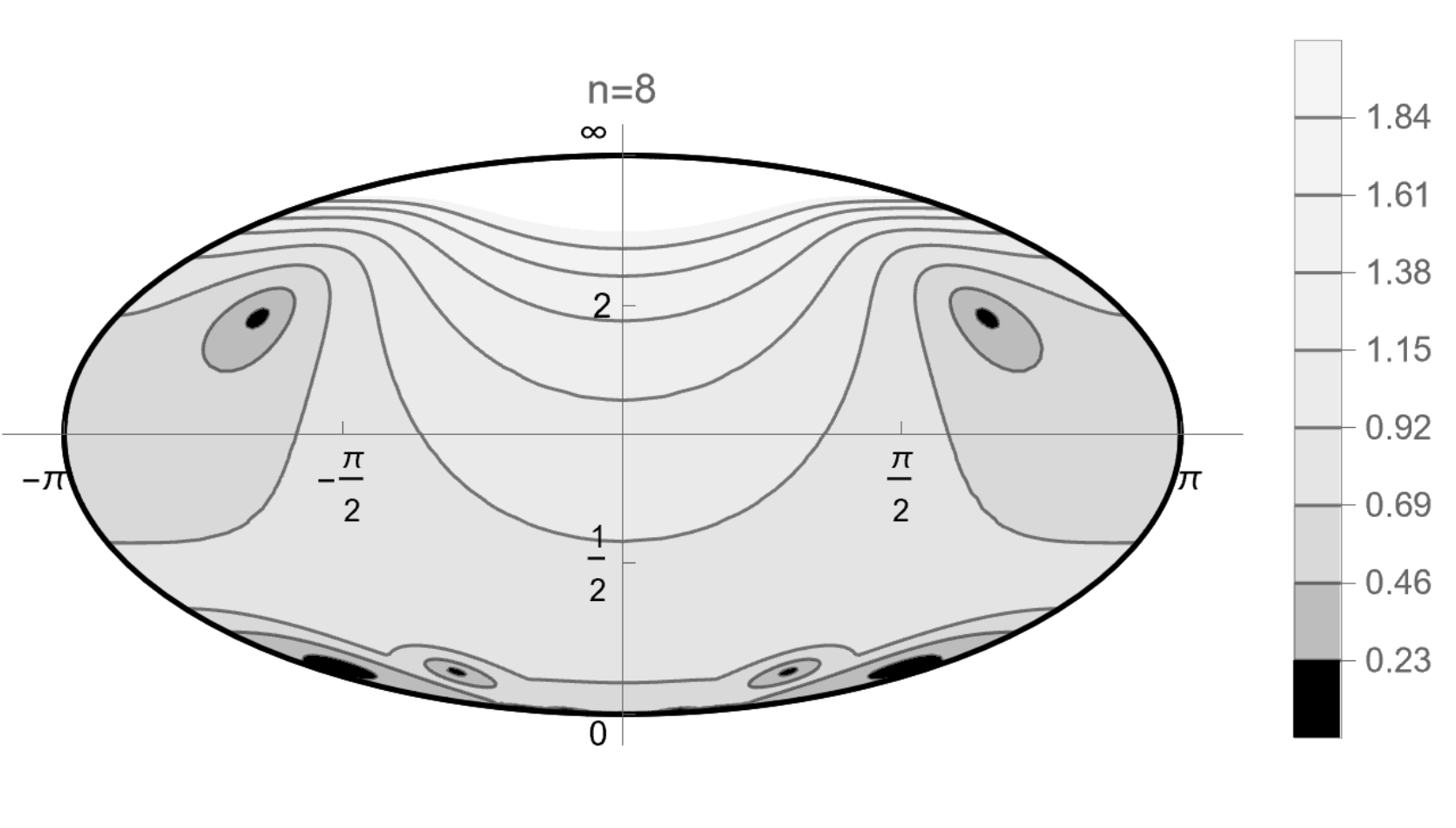}

\includegraphics[width=8.6cm]{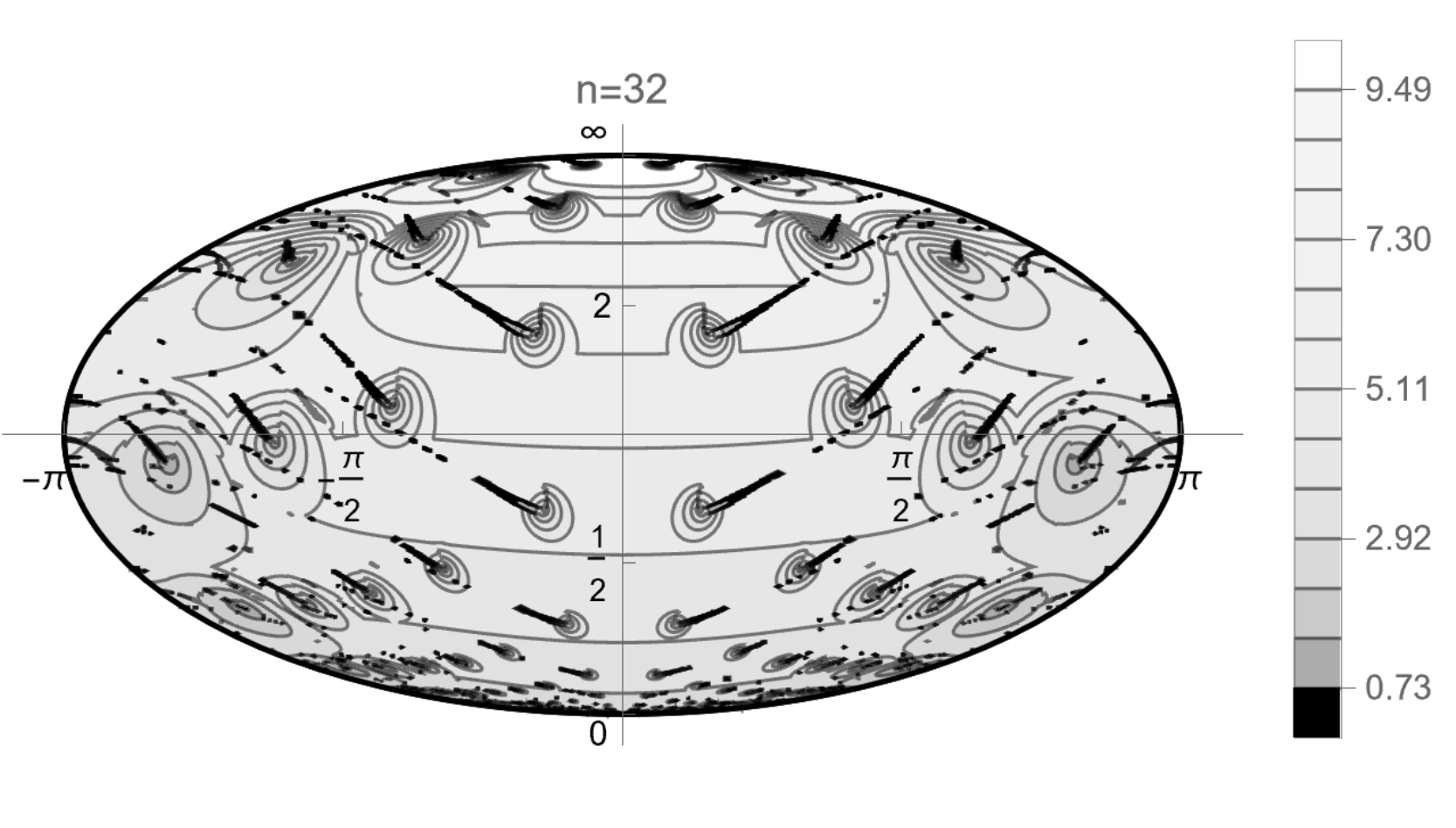} 
\caption{\label{fig:riemann} Mollweide projections of contour plots of $\min_{i,j}\left|z_i - z_j\right|$ on the Riemann sphere for $n=8$ (top) and $n=32$ (bottom). The left and right edges of each ellipsoidal outline are identified and represent the negative real axis, while the positive real axis runs -- apparently without obstruction -- down the middle.  }
\end{figure}

The conclusion of this section is that in the three cases considered, there is a disk in the complex $\lambda$ plane where all the perturbative series converge. In addition, analytical continuations in the positive real direction seem possible. As shown in 
Fig. \ref{fig:riemann} for $\nmax$ = 8 and 32, the singularities do not seem to pinch the positive real axis and it seems possible to cover the entire positive real axis by analytic continuation. The Mollweide projection \cite{moll} where the positive real axis is mapped into the central vertical line was used.

It seems plausible that for the single site Hamiltonian in general, there exists a finite radius of convergence 
for the weak and strong coupling expansions and no singular points pinching the positive real axis.
This suggests that analytic continuation could be carried out between 0 and $+\infty$ along the positive real axis (or between the north and south poles on the Riemann sphere) at every even $\nmax$, perhaps allowing a meaningful limit to be taken as $\nmax\to\infty$.


\section{$\phi^4$ in 1+1 dimensions}
\label{sec:2d}
In this section, we consider $N_s$ coupled oscillators with a given $\nmax$ in one spatial dimension. The dimension of the Hilbert space is $\nmax^{N_s}$ and the symbolic calculation of the determinant appearing in the characteristic equation becomes rapidly impossible as we increase $N_s$. In addition, when the hopping term $\kappa$ introduced in Eq. (\ref{eq:genham}), which couples neighboring fields, is set to zero, the spectrum 
becomes degenerate and the method to identify singularities needs to be reconsidered. Nevertheless, 
the calculation of the second derivative of the energy levels on the real $\lambda$ axis seems to remain a robust way to estimate the radius of convergence, even at higher energy levels. Though the size of the operator $H$ grows exponentially 
($\nmax^{2 N_s}$), the actual number of nonzero matrix elements grows at a significantly slower rate (~$\left(\nmax^{1.3 N_s}\right)$), making this a problem well-suited to the use of sparse GPU algorithms. In particular, if only the few smallest eigenvalues are desired, algorithms such as the thick-restart Lanczos method \cite{Wu:2000thick} can be used to extract them without diagonalizing the entire matrix.

\begin{figure}[h]
\includegraphics[width=8.6cm]{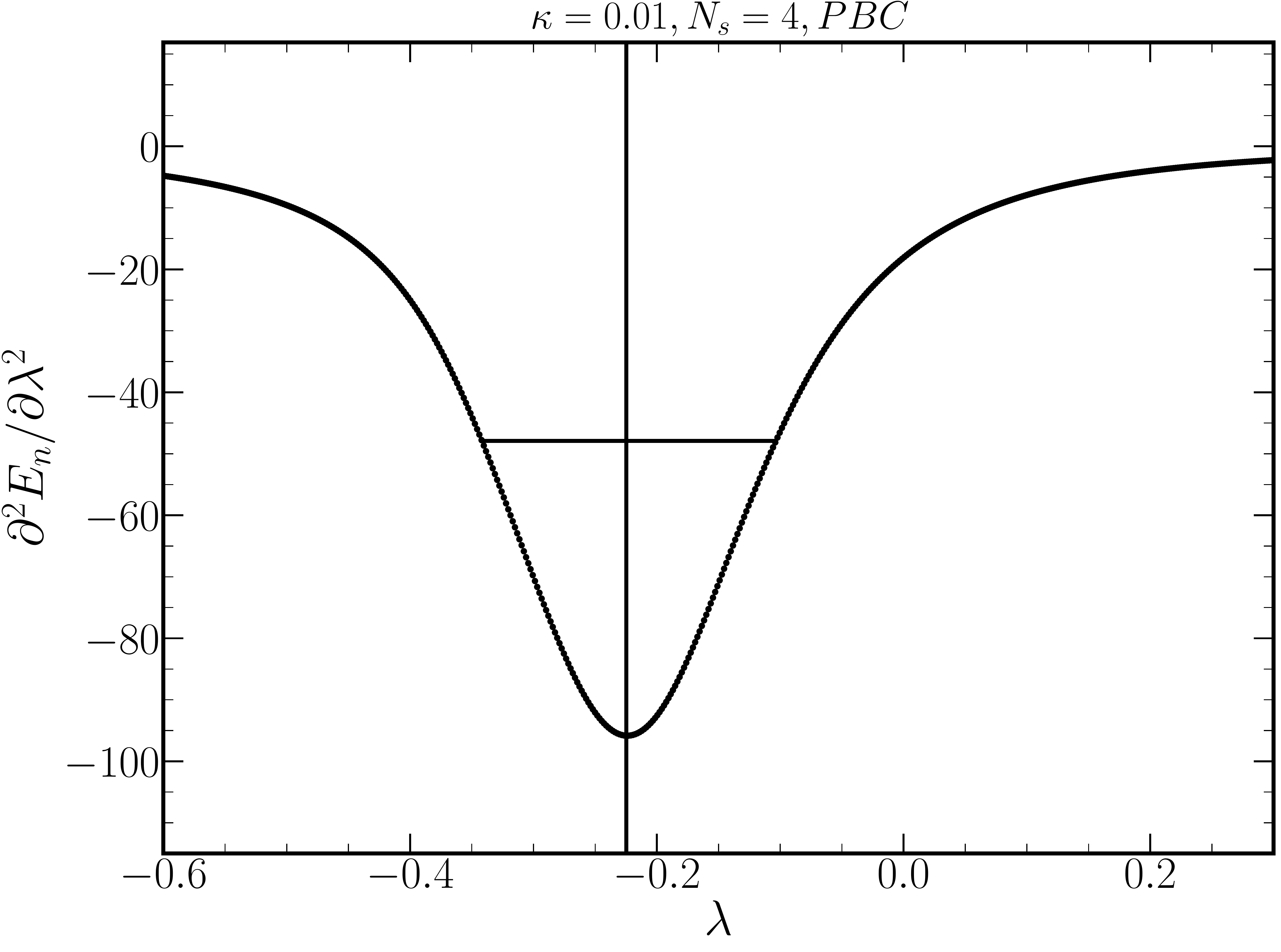} 
\vskip5pt
\includegraphics[width=8.6cm]{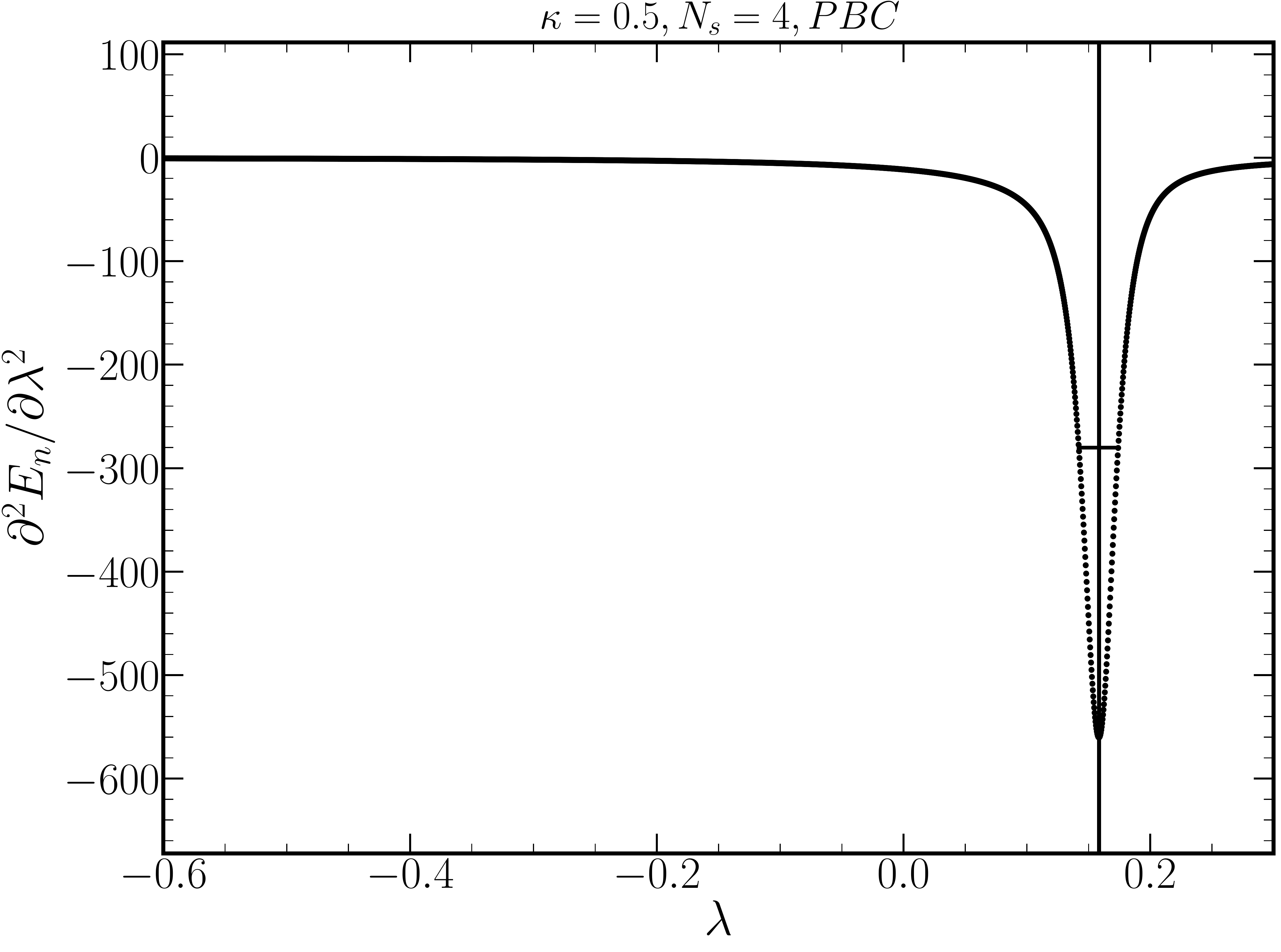} 
\caption{\label{fig:secder05}Second derivatives of $E_0$ with respect to $\lambda$ on the real axis for $\nmax=$4 and $N_s=4$, for $\kappa=0.1$ (top) and  $\kappa=0.5$ (bottom). The vertical line is located at the minima and the horizontal line represents the width at mid depth. PBC means that energies were calculated under the assumption of periodic boundary conditions on the spatial lattice.}
\end{figure}

\begin{figure}[h]
\includegraphics[width=8.6cm]{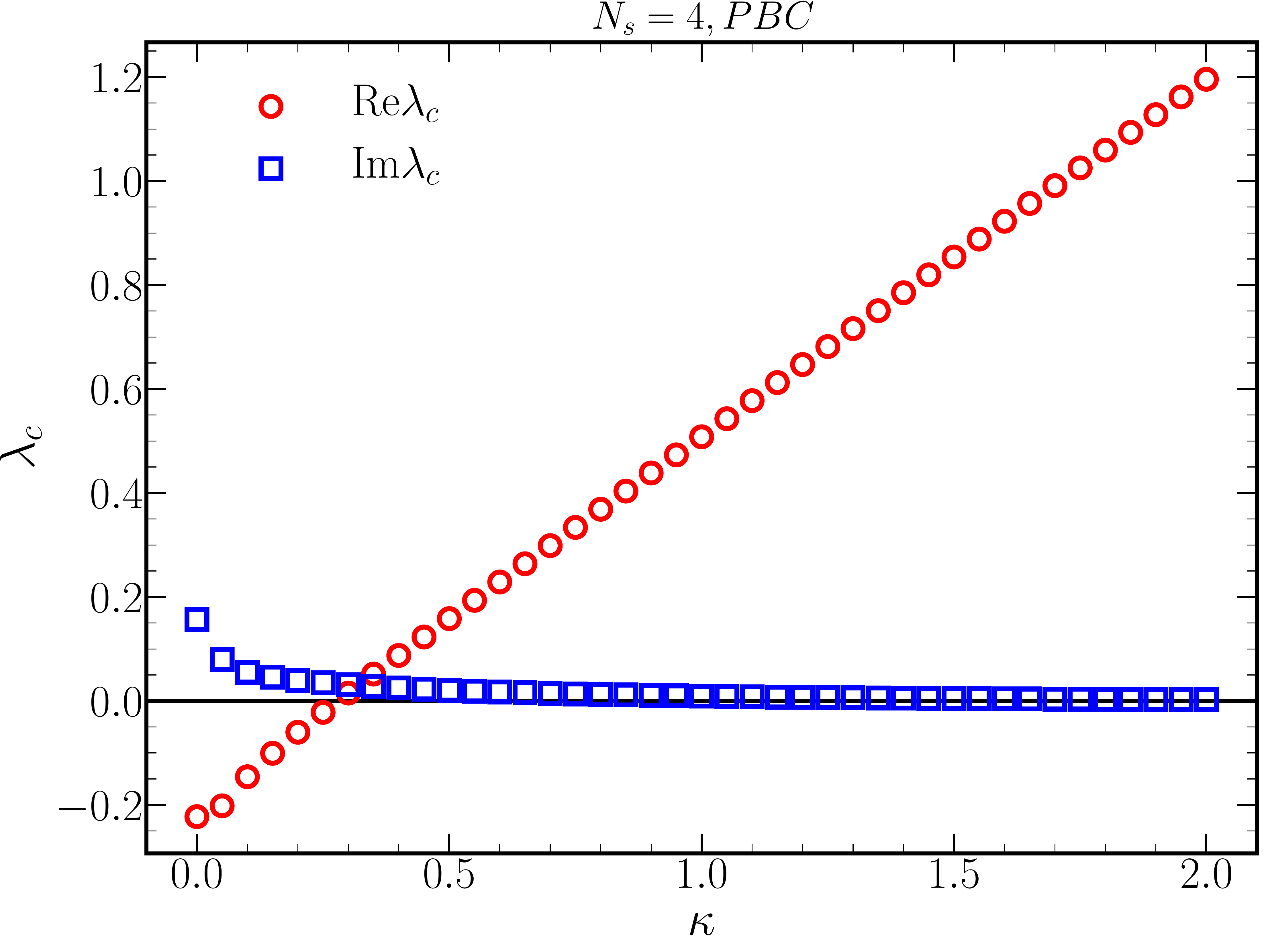} 

\includegraphics[width=8.6cm]{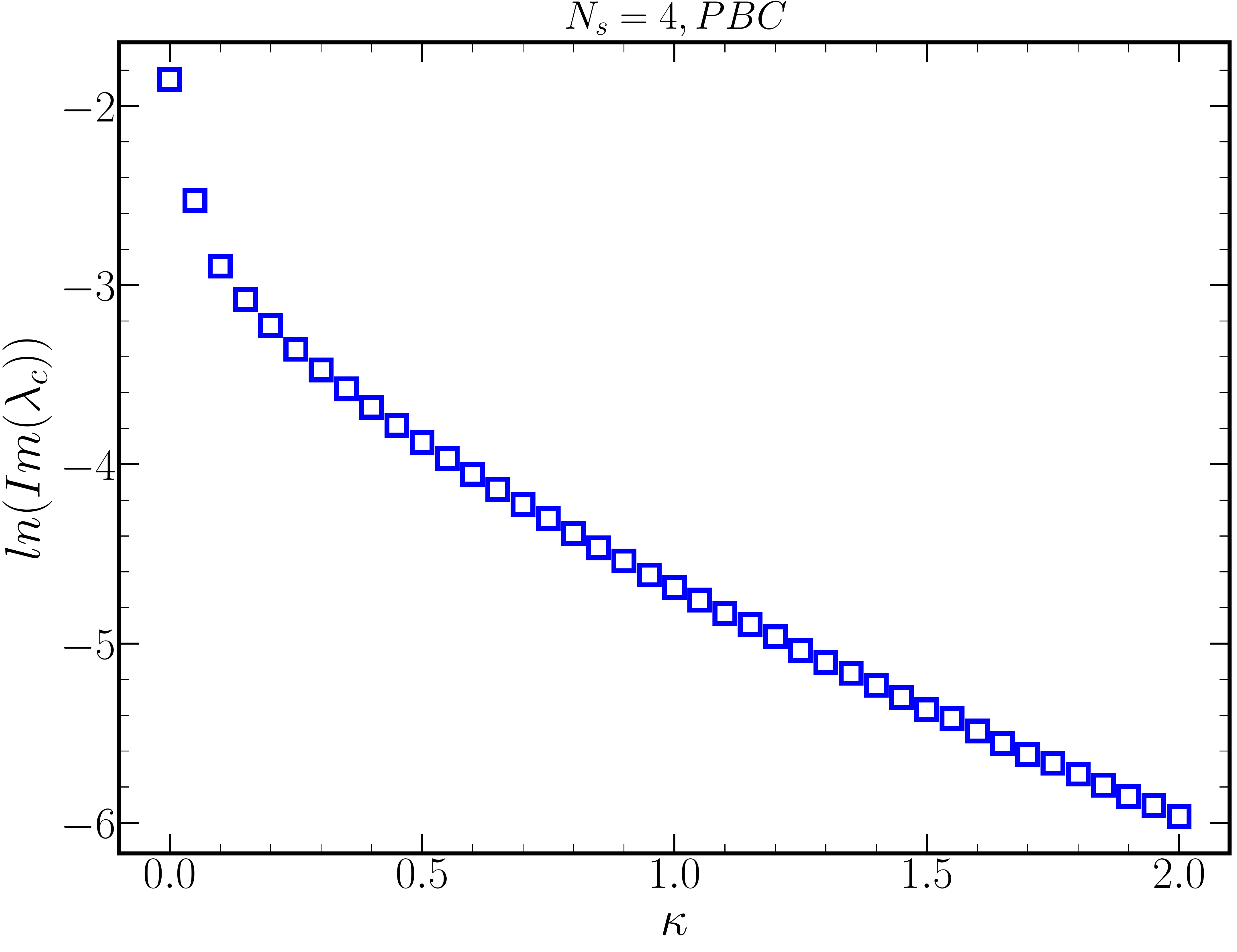} 
\caption{\label{fig:1p1} Estimates of the real and imaginary part of the singularity for $N_s=4$ $\nmax$ = 4 and different values of $\kappa$. }
\end{figure}

For small $\kappa$, the minimum appears at values close to the single site value as illustrated in Fig. \ref{fig:secder05}. As we increase $\kappa$, 
the location of the minimum increases and the width decreases as shown for $\kappa=0.5$. Estimates of the real and imaginary part of the singularity using the method developed for the single site are shown in Fig. \ref{fig:1p1}. The imaginary part seems to decreases monotonically as $\kappa$ increases. If in the large volume limit, the imaginary becomes zero for a set of values of $\kappa$, 
this would signal a phase transition which is expected to be of the Ising universality class.
Nevertheless, the perturbative series could have a finite radius of convergence on both sides.

\section{Universal quantum computing}
\label{sec:qc}
In this section, we discuss the practical steps to implement the real-time evolution described above with a universal quantum computer.
From recent work on this question \cite{jordan2012quantum,somma2016quantum,macridin2018digital,klco2018digitization,Barata:2020jtq,Kurkcuoglu:2021dnw}, it is clear that 
for an ideal universal quantum computer with many qubits, the time evolution can be calculated efficiently. In the following, we focus on practical implementations with currently available NISQ devices and the presentation is focused on one site with $\nmax$ = 4 and $n_q$ = 2, but this can be extended easily for larger systems. 

\subsection{The quantum circuit}
\begin{figure}
\includegraphics[width=14cm]{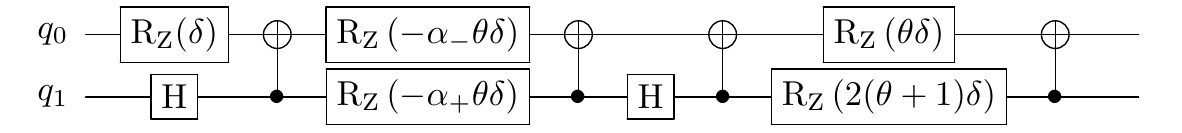}
\caption{One Trotter step in the evaluation of $\hanh$ at $\omega=1$ and  $n_\mathrm{max}=4$ (two qubits). The Trotter step size is given by $\delta$, and the following abbreviations are used for brevity: $\alpha_\pm = \sqrt{2} \pm \sqrt{6}$, $\theta=\frac{-3\lambda}{2}$.}
   \label{fig:qcirc}
\end{figure}

Following \cite{klco2018digitization}, we can express the Hamiltonian by using a representation of the operators in a tensor product basis where  elements of the Pauli group act on each of the qubits. 
The Pauli group $\left\{\bigotimes_i^{n_q} \sigma_{a_i} \middle| \forall i, a_i\in[0,3]\right\}$ of $n_q$-fold tensor products of Pauli matrices forms an orthonormal basis for Hermitian matrices under the inner product defined by $\left\langle A,B\right\rangle = \mathrm{tr}\left(B^\dagger A\right)$. This permits a Trotter reduction of an unitary $2^{n_q} \times 2^{n_q}$ matrix $U$  written in the form $e^{iG}$ by reducing the Hermitian matrix $G$ into a sum of Pauli gates and then using the Suzuki-Trotter expansion to write the resulting exponential into a power of products of coupled rotation gates. For example, $\hanh$ for $\nmax$ = 4 given in Eq. (\ref{eq:hanh}) can be broken down into the sum
\begin{align}
\hanh &= \left(\frac{15}{4}\lambda + 2\right) \mathbbm{1} \otimes \mathbbm{1} + \frac{3\lambda}{4}\left(\sqrt{2}+\sqrt{6}\right) \sigma_x \otimes  \mathbbm{1} \nonumber\\
& - \left(\frac{3\lambda}{2} + 1\right) \sigma_z \otimes \mathbbm{1} - \frac{1}{2} \mathbbm{1} \otimes \sigma_z - \frac{3\lambda}{2} \sigma_z \otimes \sigma_z\nonumber\\
& + \frac{3\lambda}{4}\left(\sqrt{2}-\sqrt{6}\right) \sigma_x\otimes\sigma_z.
\end{align}
Upon exponentiation via the simplest Trotter approximation, it can be compiled into the circuit depicted in Fig. \ref{fig:qcirc}. Examples of related circuits with $n_q \geq 3$ are depicted in Ref. \cite{klco2018digitization}.

\begin{figure}
    \centering
    \includegraphics[width=8.cm]{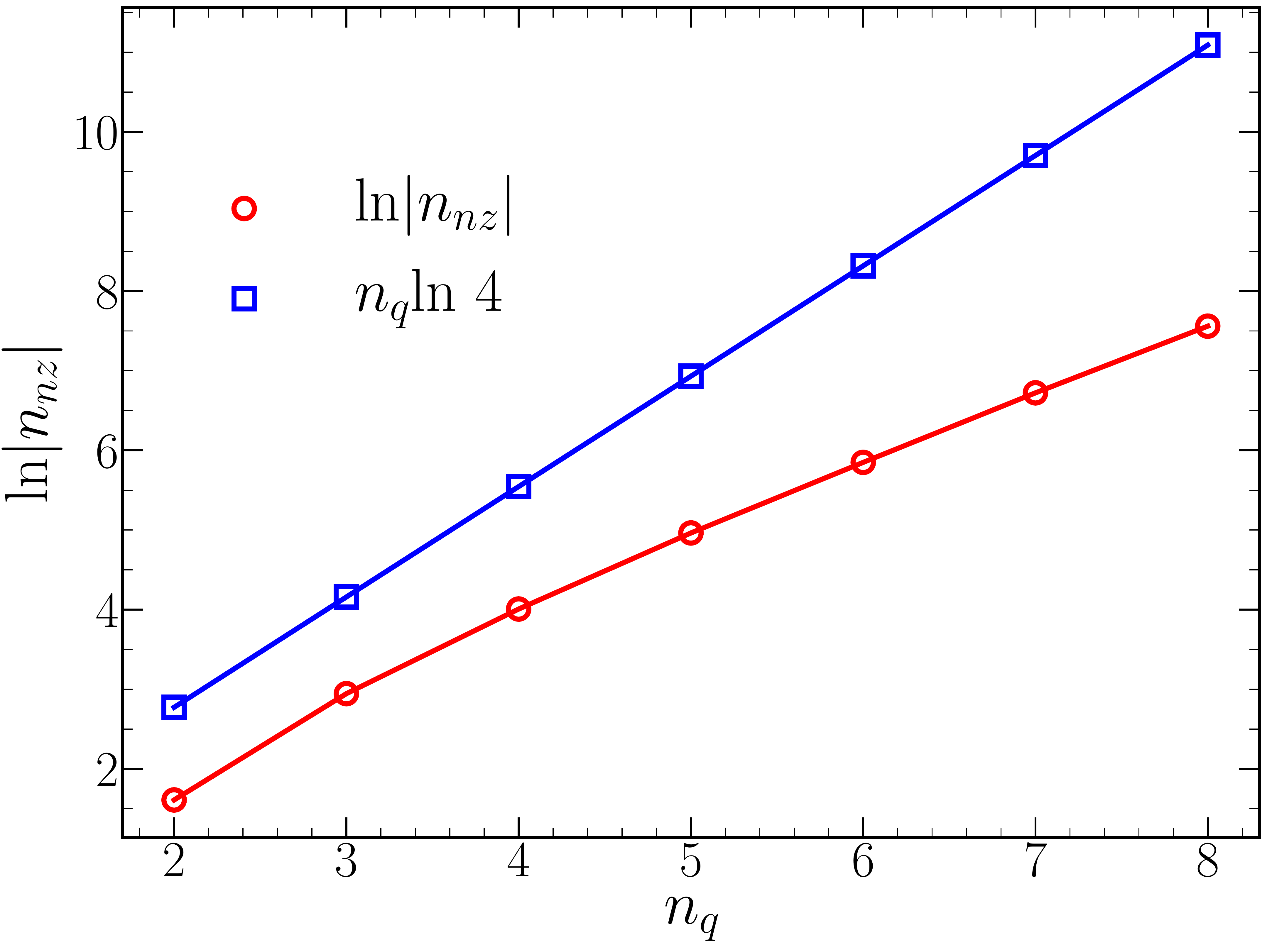}
            \caption{Plot of $\ln n_{\mathrm{nz}}$, the logarithm of  the number of nonzero components in the expansion of $\hanh$ into Pauli gates.}
    \label{fig:lnnnz}
\end{figure}

Since the number of basis operators increases exponentially (as $4^{n_q}$, with $n_q = \lg \nmax$ the number of qubits), it is reasonable to ask if $n_{\mathrm{nz}}$, the number of nonzero components in the expansion of $\hanh$ into Pauli gates does as well. The numerical results---neglecting the identity term because it exponentiates into an unobservable global phase---are shown in Table \ref{tab:nz}.
Fig.  \ref{fig:lnnnz} indicates that $n_{\mathrm{nz}}$ grows slower than $4^{n_q}$.
\begin{table}[h]
 \begin{tabular}{|c|c c c c c c c|}
 \hline
 $n_q$&2&3&4&5&6&7&8\\
    \hline
    $n_\mathrm{nz} $& 5 & 19 & 55 & 143 & 347 & 831 & 1920 \\
    \hline
    $\lceil N_{\phi^4} \rceil$ & 25 & 133 & 495 & 1573 & 4511 & 12465 & 32640\\
    \hline
   \hline
\end{tabular}
\caption{Resources required to implement $\hanh$ at $n_q$ from $2$ to $8$ ($\nmax$ from $4$ to $128$). $\left\lceil N_{\phi^4} \right\rceil$ provides a pessimistic upper bound on the number of gates required to implement $\hanh$ on a universal quantum computer. \label{tab:nz} }
\end{table}

 We may then estimate $N_{\phi^4}$ -- the compiled gate depth, the largest number of elementary gates actually implemented on hardware on any one qubit by the circuit -- by noting that the coupled phase gate on $n_q$ qubits $\sigma_z^{\otimes n_q}$ requires $n_q - 1$ pairs of CNOT gates, plus a single-qubit phase gate. Turning such a gate into any other rotation gate with the same arity -- that operates on the same number of qubits -- requires an additional pair of basis-change gates per qubit, all of which however can be done in parallel as long as $W=\sqrt{X}$ is implemented on hardware. Thus, the gate depth of a single coupled-rotation gate is at most $2 n_q + 1$, and
\begin{align}
N_{\phi^4} &\leq n_\mathrm{nz} (2 n_q + 1)
\end{align}
(We indicate that this is a pessimistic upper bound by the notation $\left\lceil N_{\phi^4} \right \rceil$, for example in Table III.)

This is estimated using a pessimistic assumption of fundamental complexity, where the average gate depth per logical coupled-rotation gate cannot be meaningfully reduced below $2 n_q + 1$. However, as demonstrated in Fig. \ref{fig:qcirc} (which has a gate depth of $8$, compared to an unoptimized depth of $5\times 5 = 25$), dramatic savings in gate depth are often possible as a result of creative reordering and optimization, so the true gate depth is likely to differ significantly from the upper bound in practice. In any case this only brings a linear factor in $n_q$ compared to $n_\mathrm{nz}$.

\subsection{Circuit execution}
\begin{figure}
    \centering
    \includegraphics[width=8cm]{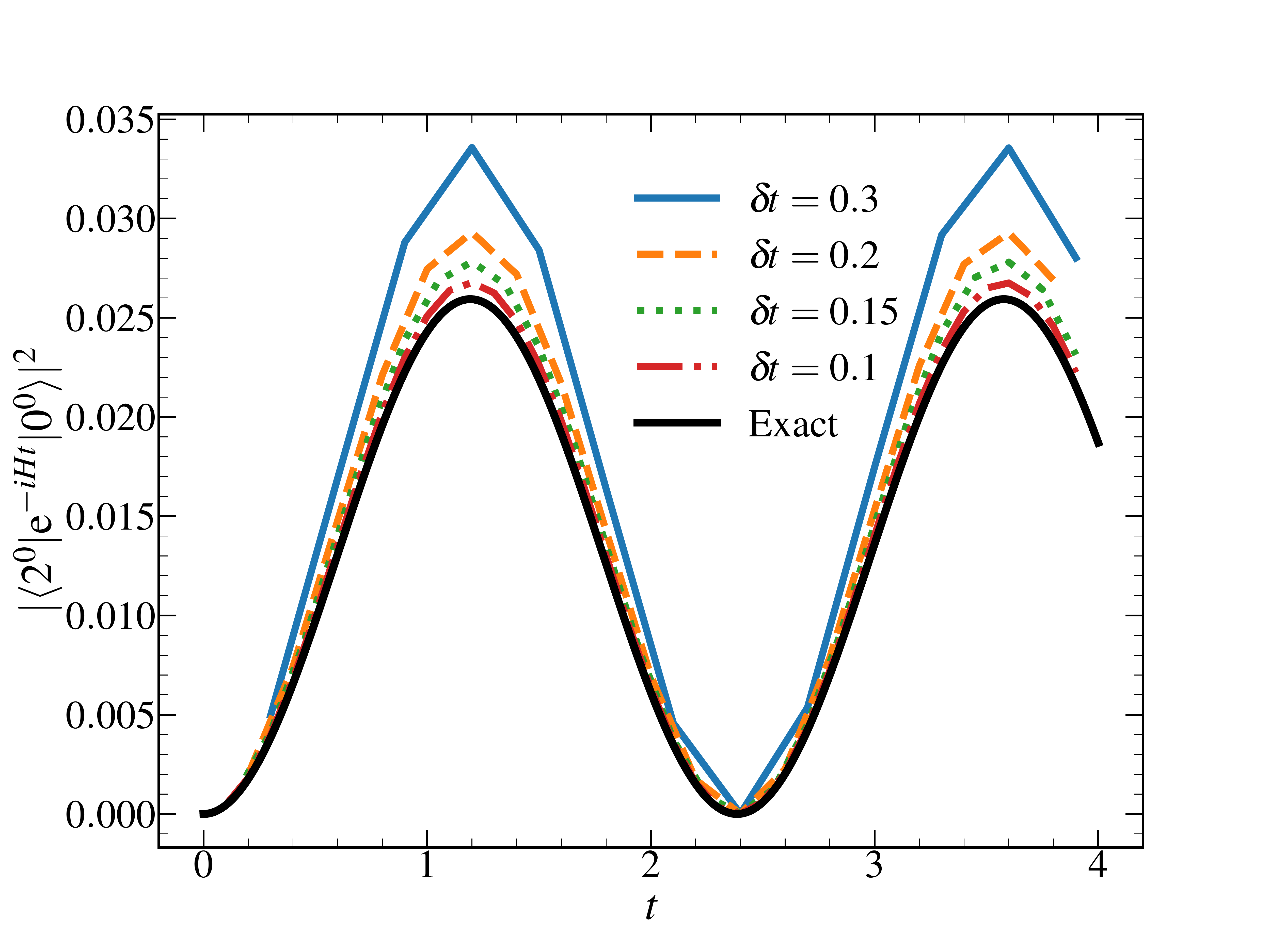}
    \caption{Time evolution of the unperturbed second excited state when the system begins in the unperturbed ground state. The dashed line is calculated exactly; the other three lines, top to bottom, are using the quantum circuit from Fig. \ref{fig:qcirc} with Trotter step $\delta t = 0.2, 0.15, 0.1$, as calculated on an ideal quantum computer.}
    \label{fig:qc_evo}
\end{figure}

Now that we have a concrete circuit to try, we can perform time evolution simulations. We will first ignore the noise from the circuit itself, and consider only the error due to using the Trotterized circuit as an approximation to the true exponential operator. Since we are simulating the smallest meaningful model at $\nmax=4$, we imagine an equally minimal computer, with two communicating qubits, a universal operation set ($\mathrm{\sqrt{X}}$, $\mathrm{H}$, $\mathrm{CNOT}$, and $\mathrm{R_Z}$) and zero gate noise. A plot of the resulting simulations at various Trotter step sizes can be found in Fig. \ref{fig:qc_evo}. A time step $\delta t$ of $0.1$ gives a graph very close to the exact result, but at the large cost of 80 gate depth per simulation time unit; if moderate errors (of up to 10\% in magnitude) are acceptable, that number can be halved, though it is still rather high in the context of existing computers. Larger steps than depicted are largely only useful for estimating frequency, as they quickly become unstable. Note that the Trotter errors in Fig. \ref{fig:qc_evo} can be compared with the perturbative errors in Fig. \ref{fig:succ} and that approximate correspondences can be drawn in that they both become larger when $\lambda$ increases.

We have started experimenting with IBMQ Qiskit simulators and public devices. For $\delta t$ = 0.2, it is possible to carry simulations for eight steps with reasonable accuracy assuming a noise model corresponding to a Falcon 8. When run on a real machine -- specifically, IBM Manila, a Falcon r5.11L machine featuring a simple chain of five qubits -- the number of steps corresponding to reasonable accuracy appears to be much shorter. After only five steps, more than a sixth (17.5\%) of trials starting in the ground state ended up in parity-forbidden odd states.
\section{Conclusions}
\label{sec:conclusions}

In conclusion, we have provided numerical evidence that the process of digitization of lattice $\lambda \phi^4$ theories results in finite nonzero radius of convergence for weak and strong coupling expansions. For quantum mechanics (single site anharmonic oscillator), analytical continuation appears to be possible on the positive real axis. For field theory in 1+1 dimensions, it seems plausible that singularities pinch the positive real axis and that separate expansions converge in the two phases.
The radius of convergence shrinks as $\nmax$ increases and the results described here suggest that one should first focus on $\nq=2$ and 3. The Trotter and perturbative errors both increase with $\lambda$ and with the current NISQ universal computers, the two approaches (perturbation theory and universal quantum computing) seem quite complementary. 

So far we have explored small systems with up to eight qubits. Going beyond this scale and in particular exploring larger dimensions would require optimizations. For intermediate couplings long perturbative series are needed and their computational cost
appear to grow exponentially  with the order. Unless efficient sampling methods \cite{dmc} can be developed for the perturbative series, it seems that there is a quantum advantage for large systems at intermediate couplings. We are planning to pursue the two ``competing" directions and compare the efficiencies with optimized methods. We also plan to use Gaussian quadrature methods and the corresponding algebraic constructions for scalar models with $O(2)$ symmetry. 

\section*{Acknowledgements} This research was supported in part by the Dept. of Energy
under Award Number DE-SC0019139. We thank Erik Gustafson, Hao Fang, S. G. Rajeev, members of the Ghent U. quantum group and members of the QuLAT collaboration for valuables suggestions. We thank Kenny Heitritter and Zeineb Mezghanni for discussion about implementation of the quantum circuits and 
Weimin Han and Stephen Jordan for valuables comments on the manuscript. 

\hfill
\break
$\ $

\appendix
\section{Perturbative construction of projectors}
\label{app:projectors}
The matrix elements of $P_0\equiv\ket{0}\bra{0}$ in the unperturbed basis for $\nmax=4$  are 
\beq
\left(
\begin{array}{cccc}
 \frac{1}{64} \left(-1701 \lambda ^4+432 \lambda ^3-72 \lambda ^2+64\right) & 0 &
   \frac{3 \lambda }{8 \sqrt{2}} \left(27 \lambda ^3-27 \lambda ^2+12 \lambda -4\right)
   & 0 \\
 0 & 0 & 0 & 0 \\ \frac{3 \lambda }{8 \sqrt{2}} \left(27 \lambda ^3-27 \lambda ^2+12 \lambda -4\right)
&
   0 & \frac{9}{64} \lambda ^2 \left(189 \lambda ^2-48 \lambda +8\right) & 0 \\
 0 & 0 & 0 & 0 \\
\end{array}
\right)
\enq
and those of $P_2\equiv \ket{2}\bra{2}$
\beq
\left(
\begin{array}{cccc}
 \frac{9}{64} \lambda ^2 \left(189 \lambda ^2-48 \lambda +8\right) & 0 & \frac{3 \lambda
    \left(-27 \lambda ^3+27 \lambda ^2-12 \lambda +4\right)}{8 \sqrt{2}} & 0 \\
 0 & 0 & 0 & 0 \\
 \frac{3 \lambda  \left(-27 \lambda ^3+27 \lambda ^2-12 \lambda +4\right)}{8 \sqrt{2}} &
   0 & \frac{1}{64} \left(-1701 \lambda ^4+432 \lambda ^3-72 \lambda ^2+64\right) & 0 \\
 0 & 0 & 0 & 0 \\
\end{array}
\right)
\enq
These imply
\beq
E_0=-\frac{567 \lambda ^4}{32}+\frac{27 \lambda ^3}{4}-\frac{9 \lambda ^2}{4}+\frac{3
   \lambda }{4}+\frac{1}{2}\enq
   and

   \beq
   E_2=\frac{567 \lambda ^4}{32}-\frac{27 \lambda ^3}{4}+\frac{9 \lambda ^2}{4}+\frac{27
   \lambda }{4}+\frac{5}{2}
   \enq

   The successive approximations for the four elements relevant for the calculation of $\bra{2^0}\hat{U}(t)\ket{0^0}$ are given in Table \ref{tab:succ}.
  
  \begin{table}[h]
 \begin{tabular}{|c|c|c|c|c|c|c|}
\hline
{\rm Order}&0&1&2&3&4&$\infty$ \\
\hline
 $E_0$&0.50000 & 0.575 & 0.5525 & 0.55925 & 0.557478&0.557806 \\
 \hline
 $E_2$&2.5000 & 3.175 & 3.1975 & 3.19075 & 3.19252&3.19219 \\
 \hline
 $\bra{2^0}P_0\ket{0^0}$&0 & -0.106066 & -0.0742462 & -0.0814057 & -0.0806897 &-0.0805242\\
 \hline
 $\bra{2^0}P_2\ket{0^0}$&0 & 0.106066 & 0.0742462 & 0.0814057 & 0.0806897&0.0805242 \\
 \hline
\end{tabular}
\caption{\label{tab:succ}Successive approximations up to order 4.}
\end{table}

\newpage

\section{From strong to weak coupling}
\label{app:stw}
\begin{figure}[h]
\includegraphics[width=8.6cm]{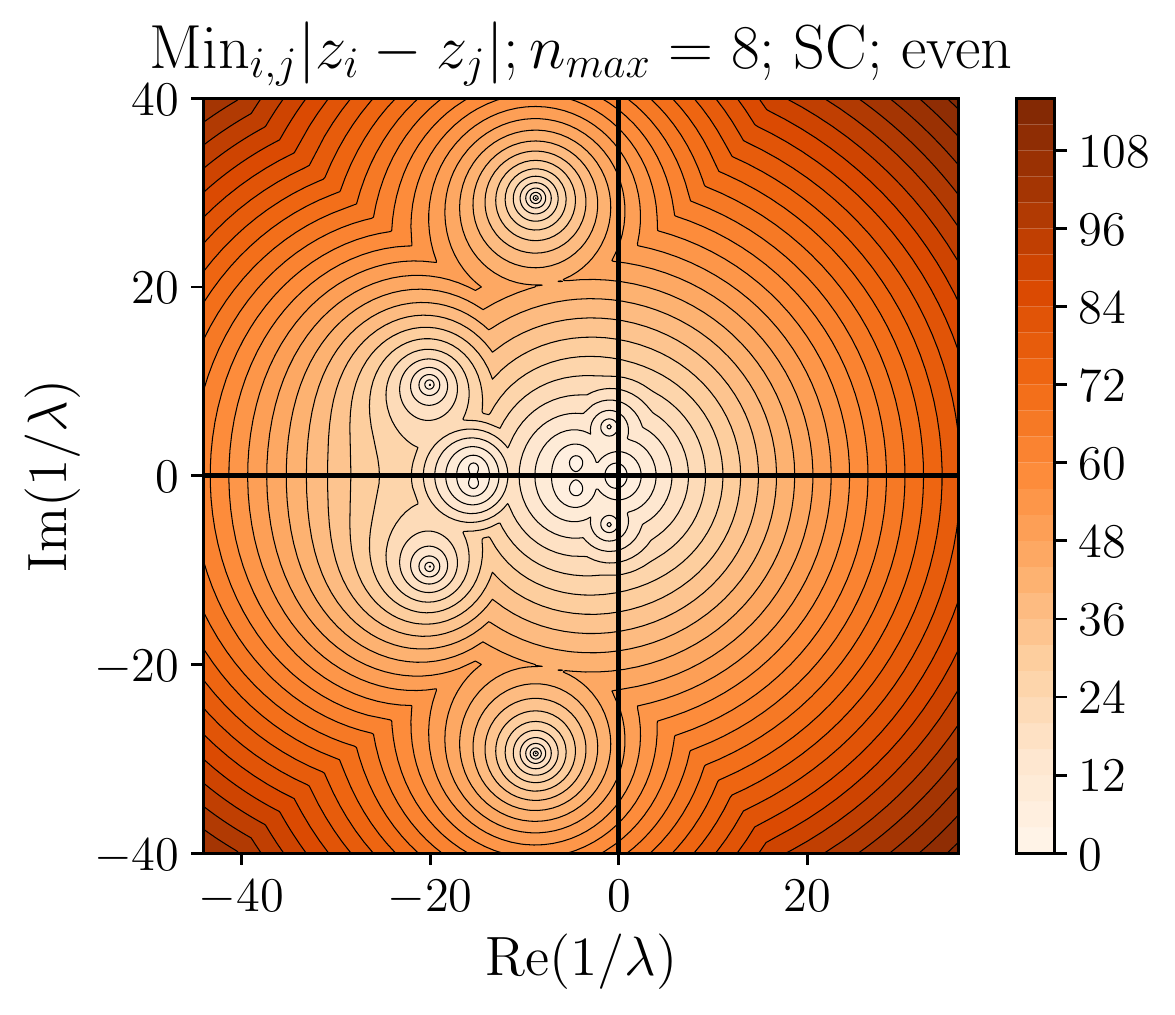} 
\includegraphics[width=8.6cm]{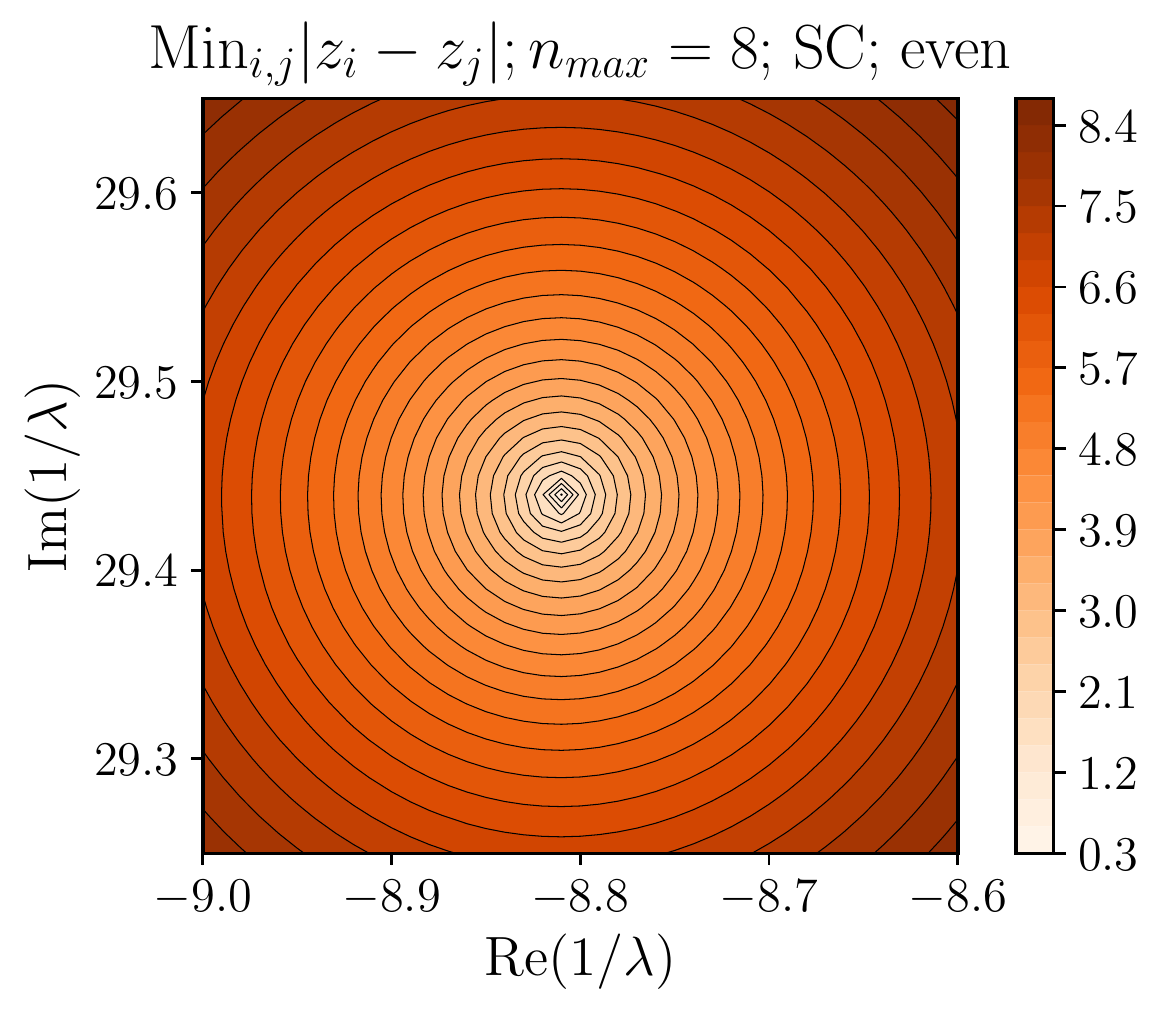} 
\caption{\label{fig:deg8evenSClarge} Minimum of $|\tilde{z}_i-\tilde{z}_j|$ for every possible pair $i,j$ of even $H^{str.}$ eigenvalues  for $\nmax$=8 in the 
complex $1/\lambda$ plane. (Top: larger scale; bottom: focus on singularity farther away from origin.) }
\end{figure}

\begin{figure}[h]
\includegraphics[width=8.6cm]{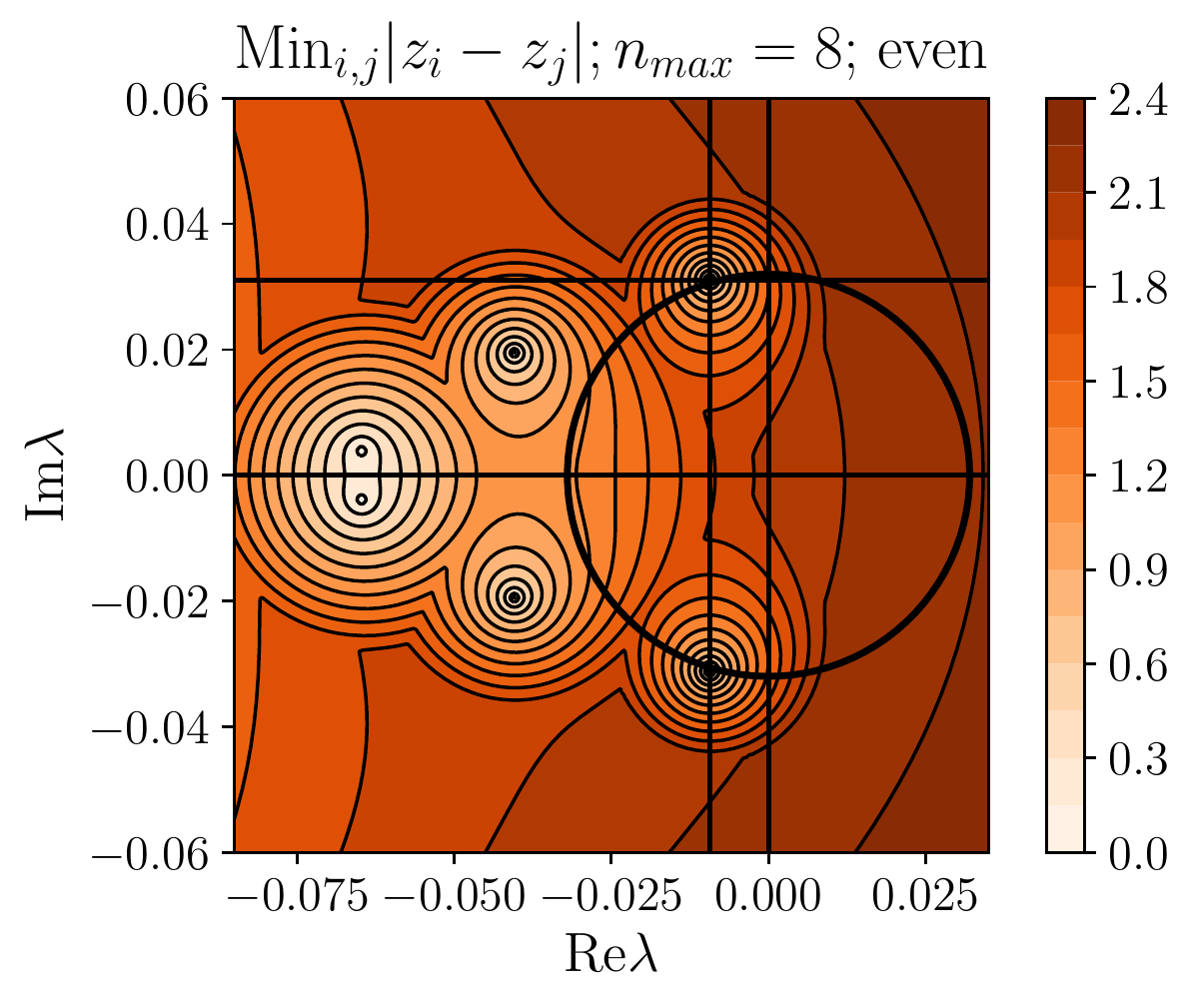} 
\caption{\label{fig:deg8evenB} Minimum of $|z_i-z_j|$ for every possible pair $i,j$ of even $H$ eigenvalues  for $\nmax$=8 in the 
complex $\lambda$ plane. 
The vertical line is located at -0.0093 and the horizontal line at 0.032.
The dark circle around the origin represents the circle of convergence of the fourth and sixth excited state and has radius 0.032.}
\end{figure}

The correspondence between the singularities in the strong vs weak coupling limit can be made precise. 
For 
instance the strong coupling singularity at approximately $-8.8+i29.45$ shown in Fig. \ref{fig:deg8evenSClarge} 
becomes $-0.0093-i0.032$ seen in Fig. \ref{fig:deg8evenB} after the $z\rightarrow 1/z$ map.

\newpage
$\ $
\newpage

\section{Algebraic equations for the singularities}
\label{app:algebra}
Double roots of the characteristic equation for $z$, when the coupling $\lambda$ is extended to the complex plane, can be found by solving simultaneously the characteristic equation $f(z,\lambda)=0$ and its derivative with respect to $z$, $f'(z,\lambda)=0$ for $z$. It can be shown \cite{hao} that this 
will yield a determinant equation for $\lambda$, that we now discuss. 
We assume that $\nmax$ is even and that the characteristic polynomial for a given parity sector has a degree $\nmax/2$.
In order to have a common root for the two equations we need to show that there is some linear relation between the two polynomials:
\beq
r(z)f(z,\lambda)+s(z)f'(z,\lambda)=0,
\enq
with $r(z)$ and $s(z)$ of degree less or equal than $\nmax/2-2$ and $\nmax/2-1$ respectively. In other words, since $f$ and $f'$ have degree $\nmax/2$ and $\nmax/2-1$ in $z$ respectively and have one common root, $f'/f$
can be written as the ratio of polynomials of degrees $\nmax/2-2$ and $\nmax/2-1$ respectively. These degrees can be reduced if additional common roots are present.
We now have to show that at least one of the following list $\mathcal{L}=z^{\nmax/2 -2}f, \dots,f,z^{\nmax/2 -1}f',\dots,f'$ is a linear combination of the others. 
This condition can be written as a determinant \cite{hao}. 

For instance, in the even sector for $\nmax=4$, using Eq. (\ref{eq:sec}), we have
\beq
f(z,\lambda)=16 z^2+(-120 \lambda -48) z+9 \lambda ^2+84 \lambda +20,
\enq
and 
\beq
f'(z,\lambda)=32 z-120 \lambda -48.
\enq
The linear dependence of one of the elements of $\mathcal{L}$ can be expressed  as a discriminant condition det$(D)=0$ for \cite{hao}
\beq
D=\left(
\begin{array}{ccc}
 9 \lambda ^2+84 \lambda +20 & -120 \lambda -48 & 16  \\
 -120 \lambda -48 & 32 & 0  \\
 0 & -120 \lambda -48 & 32 \\
\end{array}
\right)
\enq
This is now a polynomial equation in $\lambda$ and the singularity condition reads
\beq
{\rm det}(D)=-8192 \left(27 \lambda ^2+12 \lambda +2\right)=0,
\enq
in agreement with the discriminant of the quadratic equation (\ref{eq:discr}). It is possible to write similar equations for 
$\nmax$= 8 and 16. This results in lengthy expressions of order 12 and 56 in $\lambda$ respectively which appear to be numerically compatible with the direct search method. For $\nmax$ = 8, we have
\begin{eqnarray*}
f(z,\lambda)=& & 256 z^4+(-23296 \lambda -3584) z^3+\left(332640 \lambda ^2+201600 \lambda +16256\right)
   z^2\\&+&\left(-529200 \lambda ^3-1468320 \lambda ^2-448832 \lambda -25984\right) z\\&+&11025 \lambda ^4+629160 \lambda ^3+810936 \lambda ^2+186144 \lambda + 9360,
\end{eqnarray*}
and

\begin{eqnarray*}
{\rm det}(D)&=&1152921504606846976\bigl[
828875955639375 \lambda ^{12}+2446821666009000 \lambda ^{11}\\ 
&+&4112778331991700 \lambda ^{10}
+2315977875333360 \lambda ^9+729625498514388 \lambda ^8\\
&+&156815960599872
   \lambda ^7+23876641218976 \lambda ^6
   +2408867895168 \lambda ^5\\&+&157100611648 \lambda ^4+6739041792 \lambda ^3+194833408 \lambda ^2+3698688 \lambda +36864\bigr].
\end{eqnarray*}

%

\end{document}